\documentclass[10pt,aps,prx,twocolumn,longbibliography,floatfix,superscriptaddress]{revtex4-2}

\usepackage{amsmath,amsfonts,amssymb,mathtools,empheq} % Additional math symbols and environments
\usepackage{bm} % For bolding symbols
\usepackage{graphicx} % For figures
\usepackage{comment} % For commenting out sections
\usepackage[dvipsnames]{xcolor} % Names for colours
\usepackage[normalem]{ulem} % For underlining text elements
\usepackage{siunitx} % For formatting numbers with units
\usepackage{setspace}
\usepackage{xr-hyper} % For crossreferencing the Supplemental Material
\usepackage{orcidlink}
\usepackage{hyperref} % Adding hyperlinks to references

\newcommand{\AaltoAffiliation}{Department of Applied Physics, Aalto University School of Science, P.O. Box 15100, Aalto, FI-00076, Finland}
\newcommand{\TurkuAffiliation}{Department of Mechanical and Materials Engineering, University of Turku, FI-20014 Turku, Finland}

\begin{document}

%I think we don't need to emphasize the experiment in the title, but it should be strong in the abstract. However, we must make the title more general, when aiming to PRX. Here is one suggestion. Tell if you come up with better.
\title{Flat Bands from Diffraction in Periodic Systems}

\author{Joel Lehikoinen\,\orcidlink{0000-0001-8727-9822}}
\affiliation{\AaltoAffiliation{}}

\author{Rebecca Heilmann\,\orcidlink{0000-0003-2716-2571}}
\affiliation{\AaltoAffiliation{}}

\author{Aron J. J. Dahlberg\,\orcidlink{0009-0000-9083-1729}}
\affiliation{\AaltoAffiliation{}}

\author{Eero Härmä\,\orcidlink{0009-0000-6581-9664}}
\affiliation{\AaltoAffiliation{}}

\author{Malek Mahmoudi\,\orcidlink{0000-0002-9580-6220}}
\affiliation{\TurkuAffiliation{}}

\author{Arpan Dutta\,\orcidlink{0000-0002-0139-2611}}
\altaffiliation{Present address: Nanoscience Center and Department of Physics, University of Jyväskylä, P.O. Box 35, 40014, Jyväskylä, Finland}
\affiliation{\TurkuAffiliation{}}

\author{Konstantinos S. Daskalakis\,\orcidlink{0000-0002-3996-5219}}
\affiliation{\TurkuAffiliation{}}

\author{Päivi Törmä\,\orcidlink{0000-0003-0979-9894}}
\thanks{\href{mailto:paivi.torma@aalto.fi}{Contact author: paivi.torma@aalto.fi}}
\affiliation{\AaltoAffiliation{}}

\date{\today}

\begin{abstract}
% While not submitted to Nature, this abstract follows the Nature guideline for abstracts: https://www.nature.com/documents/nature-summary-paragraph.pdf
Periodic photonic structures enable precise control over the light--matter interaction through band structure engineering. Certain lattice geometries exhibit dispersionless flat bands, characterized by vanishing group velocity and diverging density of states, which present unique opportunities for applications such as slow light, nonlinear optical processes and controlling photoluminescence. However, thus far, flat bands have not been reported in systems where the lattice sites are radiatively coupled over a long range. Here we show that lattices consisting of superposed equispaced one dimensional chains exhibit flat bands with a purely diffractive origin, with the energies and angles of the flat bands controlled by the geometrical parameters of the lattice and the unit cell. The flat bands extend over all angles, can have linewidths on the order of a few nanometers, and are linearly polarized. We experimentally observe flat bands at predicted energies in lattices of gold nanoparticles at near-infrared frequencies using Fourier spectroscopy. Our results provide a general and efficient design strategy for lattices with flat, polarized dispersions for applications such as flat-band lasing, enhancing light--matter interaction, and controlling the emission or absorption of electromagnetic radiation over a wide spectral range. 
\end{abstract}

\maketitle

\section{Introduction}

Waves propagating in spatially periodic structures exhibit discrete band structures, with each band having its own dispersion relation $E_m(\mathbf{k})$, linking the energy of the $m$th band $E_m$ to the wavevector $\mathbf{k}$. \emph{Flat bands} are dispersionless bands for which $E_m(\mathbf{k}) = E_0$ at least over some range of the wavevector $\mathbf{k}$ (or angle of propagation). On a flat band, the group velocity vanishes and the density of states diverges. These properties pave the way for several interesting applications: stronger nonlinear optical processes~\cite{Sun2025, Ning2023}, optical signal processing components~\cite{Baba2008}, diffraction-free transport of images~\cite{Vicencio2015}, and increasing the rate of spontaneous emission~\cite{Purcell1946, munley2023visible} and absorption~\cite{choi2024nonlocal}. For overviews of recent research in the field, see reviews~\cite{leykam2018artificial, leykam2018perspective, Danieli2024}. Although many systems in fact have only quasi-flat bands ($E_m(\mathbf{k}) \approx E_0$), for the sake of brevity, hereinafter we use the term "flat bands". The first theoretical predictions of tight-binding (coupling between lattice sites limited to the nearest or a few neighbors) systems date from the 1980s, first in a quasiperiodic tiling~\cite{Kohmoto1986}, then in any frustrated two dimensional (2D) lattices~\cite{Sutherland1986}, a category that includes the extensively studied Lieb lattice~\cite{Lieb1989}. Lieb showed for the Lieb lattice, and Mielke for line graph lattices such as the kagome lattice~\cite{Mielke1991}, that these lattices have a ferromagnetic flat-band ground state. Further algorithms have been developed to construct lattices within the tight-binding assumption with flat-band states~\cite{Tasaki1998, Maimaiti2017, Maimaiti2019, Maimaiti2021}. Interest in flat bands grew rapidly after observation of strongly correlated and superconducting states in flat bands of graphene moir\'e systems~\cite{Cao2018-1, Cao2018-2, Balents2020, Kennes2021, Andrei2021, Torma2022}.

Advances in nanofabrication techniques permitted experimental realizations of photonic flat bands starting from 2014 in lattices of parallel waveguides~\cite{GuzmanSilva2014, Vicencio2015, Xia2018, Cantillano2018, Song2023, mukherjee2015observation}, coupled metallic resonators~\cite{Yang2024, Kajiwara2016, Nakata2012}, short-lived exciton--polaritons~\cite{jacqmin2014direct, baboux2016bosonic, harder2021kagome}, photonic moir\'e lattices~\cite{Yan2025, Wang2020, Jing2025, Yi2022, Ning2023, Mao2021, Luan2023}, strained photonic crystals~\cite{Barczyk2024}, photonic crystals with disorder~\cite{Qin2025}, photonic crystals exploiting rotational degrees of freedom~\cite{Xia2025} or by using metasurface-on-a-waveguide structures~\cite{amedalor2023high, nguyen2018symmetry, choi2024nonlocal, munley2023visible, Le2024, Choi2025, Eyvazi2025, Do2025, Choi2025-2, Sun2025-2}. Of these systems, lattices of waveguides, photonic moir\'e lattices, coupled metallic resonators, and exciton--polariton systems can be considered as emulators of tight-binding systems, because they restrict the effective coupling range between lattice sites through evanescent coupling (lattices of waveguides and photonic moir\'e lattices), physical couplings (coupled metallic resonators), or short lifetime of exciton--polaritons. In the metasurface--waveguide structures, the guided modes of the waveguide are coupled to the free space modes by the metasurface. The bands are flattened by (i) the refractive index contrast with the surrounding medium~\cite{Eyvazi2025, amedalor2023high}, (ii) by coupling of the odd and even guided modes of the waveguide through breaking of vertical symmetry~\cite{nguyen2018symmetry, choi2024nonlocal, munley2023visible, Le2024, Choi2025, Choi2025-2}, or (iii) by using strong coupling~\cite{Do2025}. In strained photonic crystals, a strain-induced pseudoelectromagnetic field flattens the dispersions of the Landau levels~\cite{Barczyk2024}. In photonic crystals exploiting disorder, the flat band forms from band folding caused by the superperiodicity and the emergence of multiple quasi-bound states in the continuum (qBICs).

\begin{figure*}[ht]
    \centering
    \begin{minipage}[c]{0.49\linewidth}
        \includegraphics[width=\linewidth]{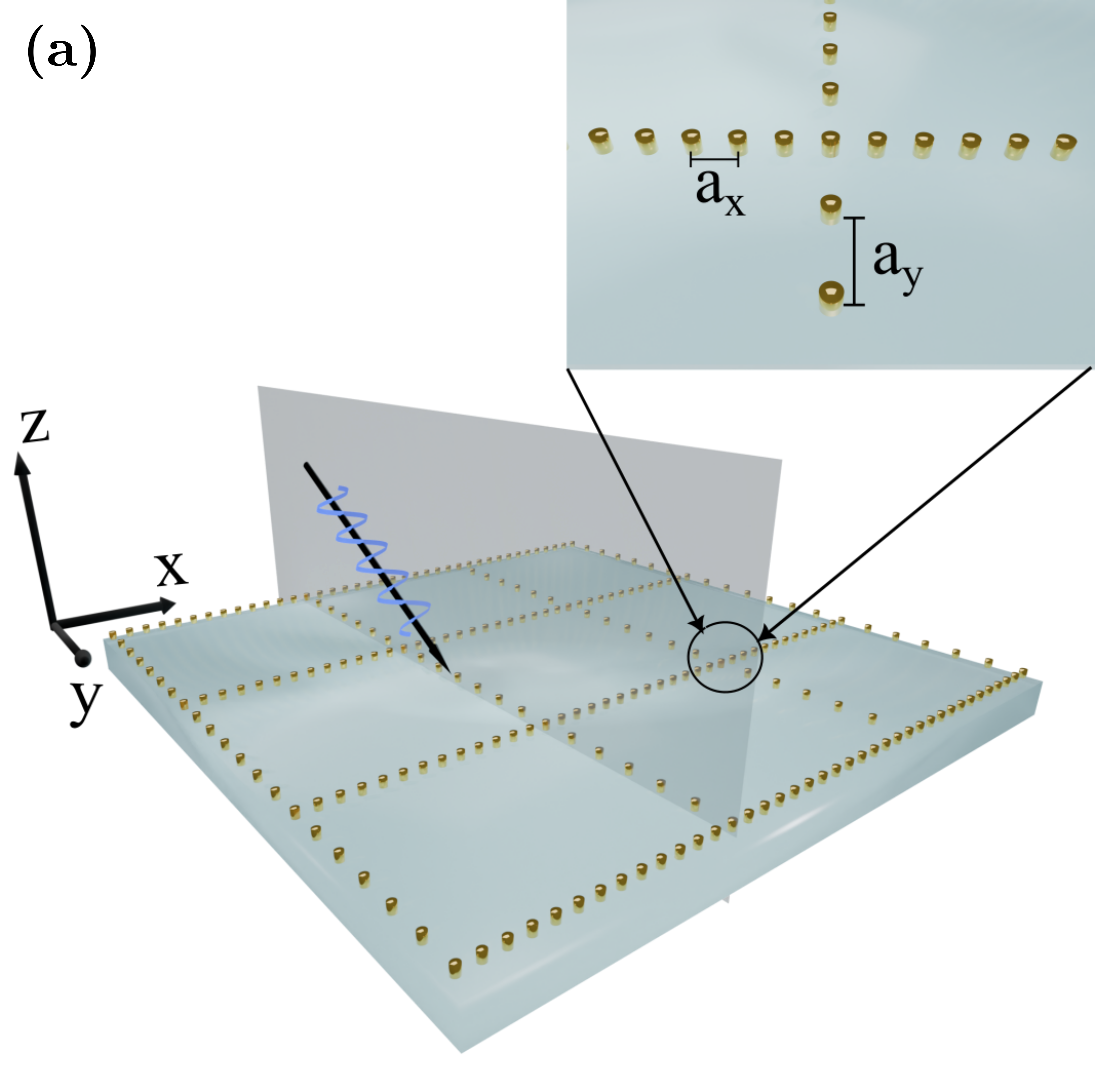}
    \end{minipage}\hfill
    \begin{minipage}[c]{0.49\linewidth}
        \includegraphics[width=\linewidth]{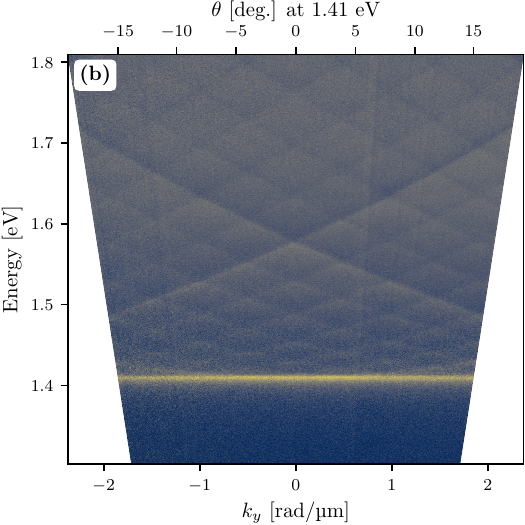}
    \end{minipage}

    \begin{minipage}[b]{0.49\linewidth}
        \centering
        \includegraphics[width=\linewidth]{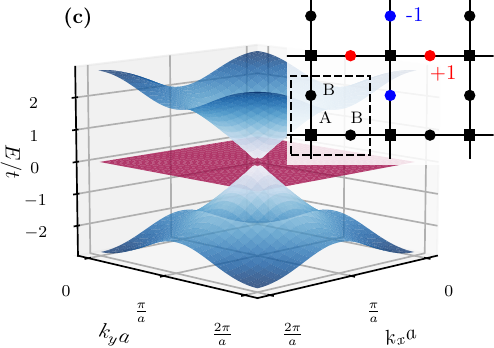}
    \end{minipage}\hfill
    \begin{minipage}[b]{0.4\linewidth}
        \includegraphics[width=\linewidth]{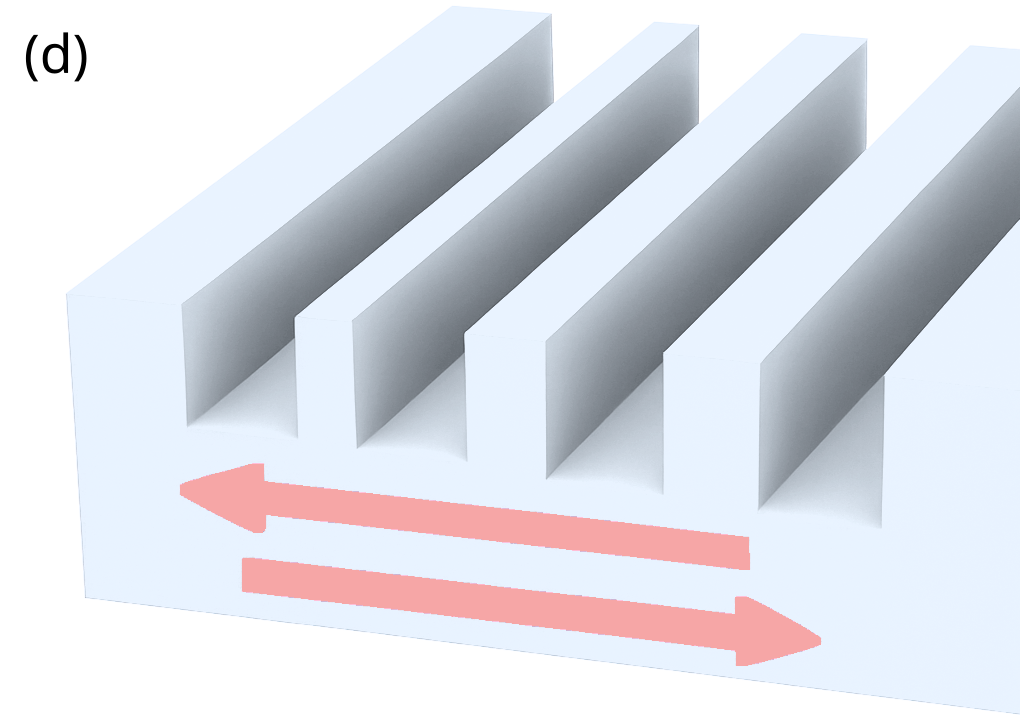}
    \end{minipage}\hspace{1cm}
    
    \caption{Schematic figure of the various means of achieving photonic flat bands. Panel (a) depicts an example of the class of flat-band lattices we propose: a rectangular chain lattice, realized here by gold nanocylinders, that diffracts incident white light preferentially into a narrow-linewidth flat band of a specified polarization. The gray plane is the plane of incidence ($yz$ plane) and here the incident light is $s$-polarized (in the $x$ direction). Panel (b) presents an experimentally observed flat band in such a geometry, discussed further in Section~\ref{s:rectangular_cross}. The in-plane momentum $k_y$ corresponds to the angle of light emission/absorption by the lattice. The color scale from blue (minimum) to yellow (maximum) shows the extinction of the array in TM polarization. Panel (c) shows a tight-binding emulating system; its top right part depicts the lattice structure of the Lieb lattice, indicating its bipartite division into $A$ (squares) and $B$ sites (circles). The dashed rectangle shows the unit cell of the lattice. The red and blue colors indicate the amplitudes of the wavefunction at the $B$ sites surrounding a single $A$ site. Hopping to the $A$ site is inhibited by destructive interference. The bottom part shows the dispersions of the three energy bands, including one exactly flat band, of the Hubbard model of the Lieb lattice. Panel (d) depicts counterpropagating guided modes (red arrows) in a metasurface--waveguide structure. The metasurface couples the odd and even modes of the waveguide, and the interference of the counterpropagating modes flattens the dispersion~\cite{munley2023visible}. In some metasurface--waveguide systems, the flattening emerges via the index contrast between the guided mode and the surrounding medium~\cite{Eyvazi2025, amedalor2023high} (not depicted here). }
    \label{f:intro_schematic}
\end{figure*}

Here, we introduce general principles to generate lattice geometries with flat bands at desired energies and angles in systems where lattice sites are radiatively coupled over a long range. The flat bands originate solely from diffraction by the lattice, and are therefore different from those described in the previous paragraph. A geometry is depicted in [Fig.~\ref{f:intro_schematic}(a)] and experimental demonstration of the concept is shown in [Fig.~\ref{f:intro_schematic}(b)]; for comparison, examples other types of flat-band generating systems are given in [Figs.~\ref{f:intro_schematic}(c)--(d)]. The flat bands generated by our approach are approximately flat ($E(\mathbf{k}) \approx \text{const.}$) at all or desired values of the wavevector $\mathbf{k}$; the flatness and angular extent are controlled by the lattice geometry, given by the general design principles expressed in analytical formulas. As the platform for experimental verification, we chose plasmonic nanoparticle lattices. Their collective optical modes, called surface lattice resonances (SLRs), are hybrids of single-nanoparticle plasmonic resonances and diffraction orders of the periodic structure~\cite{Zou2005, Auguie2008, Kravets2008, Wang2018}. The SLR modes follow quite closely the diffraction condition, and they can be approximately described by the \emph{empty lattice approximation} where the effects of a non-trivial unit cell are taken into account by a structure factor that modifies the simple diffraction condition~\cite{Guo2017}. We use the empty lattice approximation to calculate the band structures and to guide the design of the lattice geometries.

This paper is organized as follows: in Section~\ref{s:flat_bands_in_ela}, we use the rectangular lattice as an illustrative example of how flat bands are formed within the empty lattice approximation and formulate predictions of the properties of the flat bands. In Section~\ref{s:chain_lattices_main}, we present \emph{chain lattices}, a category of lattice geometries that exhibit flat bands at angles and energies controlled by the lattice geometry. Sections~\ref{s:rectangular_cross}–\ref{s:multichain_lattices} explore examples of chain lattices to study how the lattice parameters affect the band structure. We experimentally demonstrate the feasibility of the proposed concepts with plasmonic nanoparticle arrays for selected examples. As a summary, rules for designing lattices with flat bands are presented in Section~\ref{s:results_summary}. Finally, conclusions and an outlook are given in Section~\ref{s:conclusions}.

\section{Flat bands in the empty lattice approximation}
\label{s:flat_bands_in_ela}

We calculate the band structures of the investigated lattice geometries using the empty lattice approximation~\cite{Guo2017}, which consists of superposing the dispersion of the diffracted waves in a homogeneous optical background weighted with the structure factor of each mode. This approach assumes that each lattice site causes scattering of the light, and is thus general for any type of approximately pointlike scatterers. We denote the number and position of unit cells of the lattice by $N_L$ and $\mathbf{R}_n$, respectively, and the number of sites in the unit cell and their distances from $\mathbf{R}_n$ by $N_u$ and $\bm{\delta}_j$, respectively. The structure factor $S_\text{tot}(\mathbf{k})$ describes the intensity of scattering by the lattice to the mode $\mathbf{k}$ and is defined as
\begin{eqnarray}
    S_\text{tot}(\mathbf{k}) &=& \frac{1}{(N_L N_u)^2}\left\vert\sum_{n=1}^{N_L} \sum_{j=1}^{N_u} e^{i \mathbf{k} \cdot (\mathbf{R}_n + \bm{\delta}_j)}\right\vert^2 \nonumber \\
    &=&  \frac{1}{N_u^2} \left\vert \sum_{j=1}^{N_u} e^{i \mathbf{k} \cdot \bm{\delta}_j}\right\vert^2 \frac{1}{N_L^2}\left\vert\sum_{n=1}^{N_L}  e^{i \mathbf{k} \cdot \mathbf{R}_n} \right\vert^2 \nonumber \\
    &=& \frac{1}{N_u^2}\left\vert\sum_{j=1}^{N_u} e^{i \mathbf{k} \cdot \bm{\delta}_j}\right\vert^2 \sum_{\mathbf{G}} \delta(\mathbf{k} - \mathbf{G}) \nonumber \\
    &\equiv& S(\mathbf{k}) \sum_{\mathbf{G}} \delta(\mathbf{k} - \mathbf{G}),
 \label{e:def_S}
\end{eqnarray}
where $\mathbf{G}$ indicates the reciprocal lattice vectors and $\delta$ is the Kronecker delta. The second-to-last equality holds for periodic and infinite lattices, for which the structure factor $S(\mathbf{k})$ can take nonzero values only at the sites of the reciprocal lattice~\cite{Egami2003}. However, it is approximately valid also in sufficiently large finite lattices, as can be seen from the excellent match between theory and experiment in the following sections. In essence, Eq.~\eqref{e:def_S} tells that the geometric arrangement of the scatterers within the unit cell can be used for controlling the value of the structure factor, that is, the weight of the mode. The structure factor can take values between 0 and 1. In the following, we consider always the part of the structure factor related to the unit cell $S(\mathbf{k})$.

%The periodic lattice diffracts incident light; the dispersion of the wave corresponding to the diffracted order $(m_1, m_2)$ is 
%\begin{equation}
%    \label{e:dispersion_diffracted}
%    E(\mathbf{k}) = \frac{\hbar c_0}{n} \left|\mathbf{k_\parallel} + m_1 \mathbf{G_1} + m_2 \mathbf{G_2}\right|,
%\end{equation}
%where $\hbar$ is the reduced Planck constant, $c_0$ the speed of light in vacuum, $n$ the refractive index of the background medium, $\mathbf{k}$ is the wavevector of the incident light and $\mathbf{k_\parallel}$ its in-plane component, and, $\mathbf{G_1}$ and $\mathbf{G_2}$ are the reciprocal lattice vectors. 

In addition to the strength of the scattering to a particular mode, we are also interested in the polarization state of the diffracted light. If the scatterers can be taken to radiate as electric dipoles---as is the case, for example, with sub-wavelength-sized plasmonic nanoparticles---diffraction in (out) of the plane of incidence corresponds to linearly TE (TM) polarized light. See Fig.~\ref{f:intro_schematic}(c) for an example: light incident on the lattice in the $yz$ plane and $s$ polarized ($p$ polarized) will induce in-plane dipolar excitations in the $x$ ($y$) direction in the nanoparticles, leading to $x$- ($y$-)polarized diffracted light. As the in-plane momentum is determined by the plane of incidence to be along $y$, the diffracted light is TE (TM) polarized with respect to this direction~\cite{Moerland2017, Kravets2018, Wang2018}. The SLRs arising from diffraction orders in (out of) the plane of incidence are customarily correspondingly called TE (TM) modes. Unpolarized light excites both TE and TM modes.

To assign a quantitative approximant for the polarization state of the mode with the in-plane wavevector $(k_x, k_y)$ and assuming the plane of incidence to be the $yz$ plane, we define the TE polarization fraction $p_\text{TE}$ as 
\begin{equation}
    \label{e:pte_definition}
    p_\text{TE} \equiv \begin{cases} 
        1,\, \text{if} \, k_x = k_y = 0, \\
        \frac{|k_y|}{|k_y| + |k_x|}, \, \text{otherwise}.
    \end{cases}
\end{equation}
If the $xz$ plane is taken to be the plane of incidence, the $x$ and $y$ components are swapped. 

Selecting the lattice positions $\left\{\mathbf{r}_n\right\}$ to produce a given structure factor profile $S(\mathbf{k})$ (and band structure) is an ill-posed problem and cannot be solved by analytical inversion. To tackle this problem, we make ansätze to find lattices with flat bands and progressively make the lattice geometry more complex based on intuition obtained from the analytical form of the structure factor. To this end, we present an example below.

\begin{figure}[ht]
    \centering
    \begin{minipage}[t]{\linewidth}
        \includegraphics[width=\linewidth]{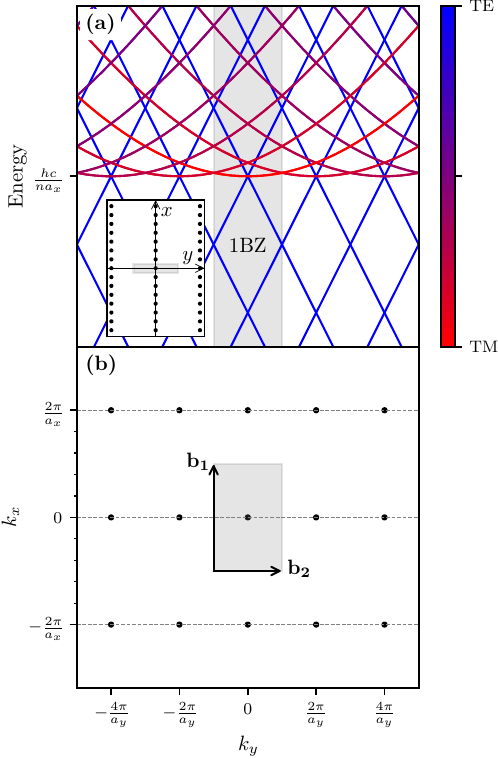}
    \end{minipage}
    \caption{Band structure (a), and the structure factor in the reciprocal lattice (b), of a rectangular lattice with $a_y = 5a_x$. (a) Flat bands form from the narrowing of the first Brillouin zone (1BZ) in the $k_y$ direction and the repetition of the mainly TM-polarized modes. At larger angles, more TE polarization mixes in. The inset shows the lattice structure and the unit cell shaded in gray. (b) The structure factor $S(\mathbf{k})$ has its maxima (black dots) at the sites of the reciprocal lattice and is zero elsewhere. The gray shaded area indicates the 1BZ; also the reciprocal lattice vectors $\mathbf{b}_1$ and $\mathbf{b}_2$ are indicated. The horizontal, gray lines in panel (b) are guides for the eye to highlight the linelike features in the structure factor. The distance between the lines is $|\mathbf{b}_1|$---this determines the energy of the flat band. Note that the axes in panel (b) are not to scale; the 1BZ is five times wider in the $k_x$ direction than in the $k_y$ direction. The minor ticks on the $k_x$ axis in panel (b) correspond to the width of the 1BZ in the $k_y$ direction.}
    \label{f:rectangular_lattice}
\end{figure}

To illustrate how the superposition of diffracting modes can form flat bands, we consider the rectangular lattice: Figure~\ref{f:rectangular_lattice}(a) shows the dispersion of a rectangular lattice, with lattice constants $a_y = 5 a_x$, along the line $k_x = 0$. A flat band is formed by the modes corresponding to the diffraction orders $(m_x=\pm 1, m_y), \, m_y \in \mathbb{Z}$, near the energy corresponding to the lattice constant $a_x$. These modes are relatively flat near their vertices and are separated in $k_y$ by the length of the reciprocal lattice vector corresponding to $a_y$. Thus, increasing the aspect ratio $a_y/a_x$ narrows the width of the Brillouin zone, yielding a flatter band. Because these modes are based on the diffractive orders $m_x=\pm 1$, they lie along the lines $k_x = \pm 2 \pi / a_x$ depicted in Fig.~\ref{f:rectangular_lattice}(b), and thus for a flat band to appear, the structure factor $S(\mathbf{k})$ must take nonzero values densely enough along these lines.

The empty lattice approximation allows us to make five predictions concerning the flat band:
\begin{enumerate}
    \item The flat band forms at an energy close to the spectral positions of the vertices of the flat-band modes, because the dispersion is the flattest near its vertex. The flat-band energy is determined by the distance of the linelike feature in the structure factor from the line of the observation (here $k_x = 0$). For the rectangular lattice, this distance is determined by the lattice constant $a_x$ via the reciprocal lattice vector $\mathbf{b}_1$ [see Fig.~\ref{f:rectangular_lattice}(b)].
    \item Because the band structure is repeated, the flat band extends over all values of $k_y$, i.e., all angles. The size of the Brillouin zone (BZ) in the $y$ direction is inversely proportional to the $y$ period. Taking it to be large, i.e., approaching the limit of a single chain of particles in the $x$ direction, leads to a vanishing BZ thus a denser set of the TM modes and thereby a flatter band. The limit of a single chain of particles in $x$ direction produces a perfectly flat band in $k_y$, consistent with the absence of any structure in the $y$ direction. However, a single chain would be inefficient in manipulating light over a large area or volume of active material.    
    \item The flat band is linearly TM polarized near $k_y = 0$; the TE polarization fraction increases away from the $\Gamma$ point.
    \item Because the linelike features formed by the maxima of the structure factor are repeated [Fig.~\ref{f:rectangular_lattice}(b)], there are flat bands at integer multiples of the energy corresponding to the lattice constant $a_x$, as will be shown in Section \ref{s:chain_lattices_main}.
    \item Its diffractive origin confers to the flat band a narrow linewidth---this is confirmed by our experiments.
\end{enumerate}

\begin{figure*}[ht]
    \centering
    \includegraphics[width=\linewidth]{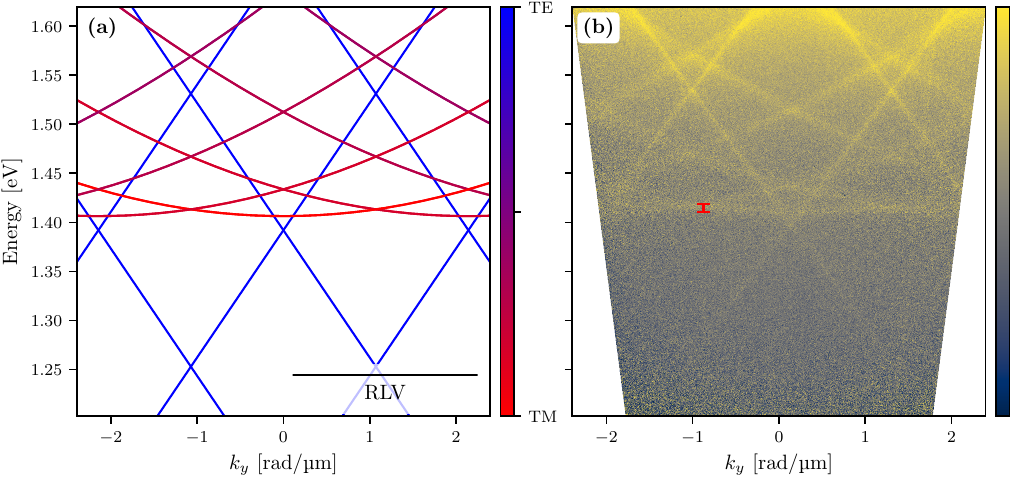}
    \caption{Calculated band structure [panel (a)] and extinction spectrum [panel (b)] of a rectangular lattice of gold nanoparticles with diameter \SI{110}{\nano\metre} and height \SI{50}{\nano\metre} with lattice constants $a_x = \SI{580}{\nano\metre}$, $a_y = \SI{2930}{\nano\metre}$ in an index-symmetric environment with refractive index $n = 1.52$. Sample fabrication (Appendix~\ref{s:sample_fabrication}) and experimental setup and procedures (Appendix~\ref{s:experimental_setup}) are described in the Appendix. The scale bar of panel (a) shows the magnitude of the reciprocal lattice vector (RLV) in the $k_y$ direction. In panel (b), the color scale from blue (minimum) to yellow (maximum) corresponds to extinction averaged over five measurements. The red bar estimates the linewidth of the flat band as \SI{8}{\milli\electronvolt} (\SI{5}{\nano\metre}).}
    \label{f:rectangular_lattice_exp_vs_ela}
\end{figure*}

To verify the feasibility of our flat-band generation strategy, we experimentally test the theoretical predictions in several example cases. The experiments are made with gold nanoparticle arrays, dimensions of the particles in the 100~nm range and of the arrays in the 200x200~square microns range. The samples are fabricated on glass substrates by electron-beam lithography (for more details, see Appendix~\ref{s:sample_fabrication}). A symmetric refractive index environment ($n=1.52$) is realized by covering the sample with index-matching oil corresponding to the glass substrate. Dispersions are measured via transmission of white light through the sample, analyzed by Fourier spectrometry, in some cases polarization filtered. More information about the experiments is given in Appendix~\ref{s:experimental_setup}. For the case of the rectangular lattice discussed above, the experimentally observed extinction spectrum and the band structure (Fig.~\ref{f:rectangular_lattice_exp_vs_ela}) calculated using the empty lattice approximation are in excellent agreement. Figure~\ref{f:rectangular_lattice_exp_vs_ela}(b) shows that the flat band extends between angles of $\pm $17\textdegree{}, corresponding to the numerical aperture (0.3) of the objective used.

In conclusion, we have demonstrated that lattices whose unit cell is much greater than the lattice constant in one direction exhibit flat bands of diffractive origin in the perpendicular direction. Furthermore, we have argued that flat bands in the band structure are equivalent to linelike features in the structure factor. This makes finding lattice geometries whose structure factors exhibit linelike features a good proxy goal in the search for lattices exhibiting flat bands. However, the simple rectangular lattice offers only two design parameters---one lattice constant to control the energy of the flat band and the other (or the aspect ratio) to control the flatness of the flat band. Moreover, the low density of lattice sites leads to weak light--matter interaction. The lattice also has a flat band in one direction only. Next, we study a category of more complex lattice geometries to alleviate these issues.

%\begin{itemize}
%    \item Introduce ELA, reference previous work on it (multimode lasing paper). Structure factor and lattices with multiple particles per unit cell. Approximations for infinite and periodic lattices.
%    \item Link between diffraction and polarisation for plasmonic NP arrays. Introduce definition for TE polarisation fraction of modes.
%    \item Brillouin zone folding in rectangular lattices, flat bands from criss-crossing TM-modes; flat band polarisation (Figure~\ref{f:rectangular_lattice}).
%    \item Analytical structure factor calculations; line-like features in structure factor imply flat bands (Figure~\ref{f:S_rectangular}); flat bands are repeated at given energies and extend over all angles.
%    \item Brief description of experimental methods; more details in the supplementary.
%\end{itemize}

\section{Chain lattices}
\label{s:chain_lattices}

\subsection{Definition and structure factor}
\label{s:chain_lattices_main}

\begin{figure}[ht]
    \centering
    \begin{minipage}[t]{0.49\linewidth}
        \includegraphics[width=\linewidth]{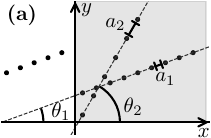}
    \end{minipage}
    \begin{minipage}[t]{0.49\linewidth}
        \includegraphics[width=\linewidth]{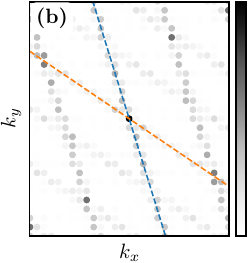}
    \end{minipage}
    \caption{Panel (a) shows a schematic of a chain lattice consisting of two chains. Lattice sites are indicated by black dots, and the unit cell of the lattice is shaded in gray. The structure factor of the lattice computed from the definition in Eq.~\eqref{e:def_S} is shown in panel (b), with darker color indicating a greater value. The blue and orange lines are the lines defined by Eq.~\eqref{e:flat_band_lines} for $l = 0$ for the angles $\theta_1$ and $\theta_2$.}
    \label{f:chain_lattice_schematic}
\end{figure}

We name as a \emph{chain lattice} any lattice whose unit cell consists of $N_c$ chains of $N_j$ lattice sites with constant spacing $a_j$ and with the chain subtending the angle $\theta_j \in [0, \pi)$ with the $x$ axis, where $j$ labels the chains. The chain is translated by $\mathbf{\Delta_j} = (\Delta x_j, \Delta y_j)$. The positions of the lattice sites of the $j$th chain are
\begin{align}
    &\biggl\{ (n a_j \cos \theta_j + \Delta x_j,  n a_j \sin \theta_j + \Delta y_j) \biggr\}_{n=0}^{N_j -1} \nonumber \\
    &\equiv \left\{n\mathbf{c_j} + \mathbf{\Delta_j}\right\}. \label{e:line_lattice}
\end{align}
Figure~\ref{f:chain_lattice_schematic}(a) shows an example of a lattice consisting of two chains. There are now $N_u = \sum_{j=1}^{N_c} N_j$ lattice sites in the unit cell; if there are any overlapping lattice sites between the chains, they must be subtracted from the number of lattice sites in the unit cell $N_u$ and the expression of the structure factor presented below. Hereinafter, we assume that no such sites exist.

The structure factor of the complete lattice is 
\begin{align}
    S(\mathbf{k}) &= \frac{1}{N_u^2} \left| \sum_{j=1}^{N_c} \sum_{n=0}^{N_j - 1} e^{i \mathbf{k} \cdot (n \mathbf{c_j} + \mathbf{\Delta}_j)} \right|^2 \nonumber \\
    &= \frac{1}{N_u^2} \left| \sum_{j=1}^{N_c} \underbrace{e^{i \mathbf{k} \cdot \mathbf{\Delta}_j}}_{\equiv \varphi_j(\mathbf{k})} \underbrace{\sum_{n=0}^{N_j - 1} \left[ e^{i \mathbf{k} \cdot \mathbf{c_j}} \right]^n}_{\equiv N_j S_j(\mathbf{k})} \right|^2. \label{e:S_chain_lattice}
\end{align}
Next, we consider the contribution of the $j$th chain $S_j(\mathbf{k})$. It takes the form of a truncated geometric series and reaches its maximum value (1) when the argument of the exponential is an integer multiple of $2 \pi i$, i.e.,
\begin{align}
    \mathbf{k} \cdot \mathbf{c_j} &= 2 \pi l, \; l \in \mathbb{Z} \nonumber \\
 \Rightarrow    k_y &= \frac{2 \pi l}{a_j \sin \theta_j} - \left(\cot \theta_j\right) k_x. \label{e:flat_band_lines}
\end{align}
This equation describes a family of lines in the $k_x$-$k_y$ plane. Because the partial sum $S_j$ is maximized along these lines, the structure factor has local maxima at the sites of the reciprocal lattice at or close to these lines. Figure~\ref{f:chain_lattice_schematic}(b) confirms that the maxima of the structure factor calculated from the definition Eq.~\eqref{e:S_chain_lattice} [or equivalently, Eq.~\eqref{e:def_S}] for a representative chain lattice appear near the lines defined by Eq.~\eqref{e:flat_band_lines}. The slope of the family of lines defined by Eq.~\eqref{e:flat_band_lines} is $-\cot \theta_j$, which corresponds to the angle
\begin{equation}
    \label{e:flat_band_angles}
    \arctan (- \cot \theta_j) = - \frac{\pi}{2} + \theta_j,
\end{equation}
that is, the flat bands can be observed at right angles to the lattice chains that generate them, consistent with the understanding given by the simple rectangular lattice example in Section \ref{s:flat_bands_in_ela}.

It is intuitively clear that the energy of the flat band corresponds to the spacing $a_j$ of the corresponding chain. Here, we show how it can be derived. As established in Section~\ref{s:flat_bands_in_ela}, in the empty lattice approximation, the flat band is observed at the energy corresponding to the $k$-space separation of a linelike feature in the structure factor and the observation line [see Fig.~\ref{f:rectangular_lattice}]. We assume the line of observation to lie along one of the linelike features in the structure factor given in Eq.~\eqref{e:flat_band_lines}. The energy of the flat bands is then given by the separation of two neighboring lines in the family of lines of Eq.~\eqref{e:flat_band_lines}. Denoting this distance by $\kappa_j$, we compute its value from the expression for the distance of two parallel lines:
\begin{equation}
    \kappa_j = \frac{2\pi}{a_j \sin \theta_j} \frac{1}{\sqrt{1 + \cot^2 \theta_j}} = \frac{2 \pi}{a_j}.
\end{equation}
Therefore, the flat bands appear at energies
\begin{equation}
    \label{e:flat_band_energy}
    E_\textrm{m} = \frac{h c_0}{n a_j} m, \; m \in \mathbb{N},
\end{equation}
where $h$ is the Planck constant, $c_0$ the speed of light in vacuum and $n$ the refractive index of the background medium. This result is evident for lattices where the particle spacing and the size of the unit cell are the same, such as the rectangular lattice of Fig.~\ref{f:rectangular_lattice}. However, in lattices with multiple lattice sites per unit cell [see Fig.~\ref{f:asymm_cross_exp_vs_sim}], the system is characterized by two distinct periodicities: that of the chain's spacing and that of the unit cell. In the following Sections, we will show that this result arises from the fact that only the diffraction orders corresponding to the chain's spacing contribute to the flat band.

%It is perhaps unexpected that the energies of the flat bands depend only on the chain spacing and not the lattice vectors (via the Rayleigh condition). When the number of lattice sites in the $j$th chain $N_j$ is sufficiently large, the chain's contribution to the structure factor (including the normalization factor $N_j^{-1}$) $S_j(\mathbf{k})$ [see Eq.~\eqref{e:S_chain_lattice}] is small for all values of the wavevector $\mathbf{k}$ except those given by the condition in Eq.~\eqref{e:flat_band_lines}. In other words, modes that do not meet the condition of Eq.~\eqref{e:flat_band_lines} interfere destructively. 

Mathematically speaking, in the limit of large $N_j$, $S_j$ becomes the Kronecker comb (sum of equispaced delta functions). In Section~\ref{s:flat_bands_in_ela}, we established creating linelike features in the structure factor as a proxy goal for attaining flat bands. Ideally, the structure factor would take the form of a Kronecker comb. On the other hand, the form of $S_j$ is essentially that of the inverse discrete Fourier transform. The discrete Fourier transform dual of the Kronecker comb is an equispaced sequence, i.e., the chain lattice of Eq.~\eqref{e:line_lattice}. Hence, we conclude that the equispaced chains are a natural way to generate linelike features in the structure factor and flat-band scattering within the purview of the empty lattice approximation.

\subsection{Rectangular chain lattices}
\label{s:rectangular_cross}

As the simplest realization of the chain lattice concept, we now consider chain lattices that consist of two perpendicular chains. Hereafter, we refer to such lattices as \emph{rectangular chain lattices}, and introduce the shorthand notation $N_x\times N_y$, where the first (second) number indicates the number of lattice sites per unit cell in the chain parallel to the $x$ axis ($y$ axis). The lattice constant of the $x$-oriented ($y$-oriented) chain is $a_x$ ($a_y$). There are $N_u = N_x + N_y - 1$ lattice sites in the unit cell; the $-1$ is to avoid calculating the lattice site at the origin twice. The lattice vectors and the reciprocal lattice vectors are
\begin{align}
    \mathbf{a}_1 &= \begin{pmatrix}
        N_x a_x \\ 0
    \end{pmatrix},
    &\mathbf{a}_2 = \begin{pmatrix}
        0 \\ N_y a_y
    \end{pmatrix}, \nonumber \\
    \mathbf{b}_1 &= \begin{pmatrix}
        \frac{2 \pi}{N_x a_x} \\ 0
    \end{pmatrix},
    &\mathbf{b}_2 = \begin{pmatrix}
        0 \\ \frac{2 \pi}{N_y a_y} 
    \end{pmatrix}. \label{e:lattice_vectors_rect_chain}
\end{align}

\begin{figure*}[ht]
    \centering
    \includegraphics[width=\linewidth]{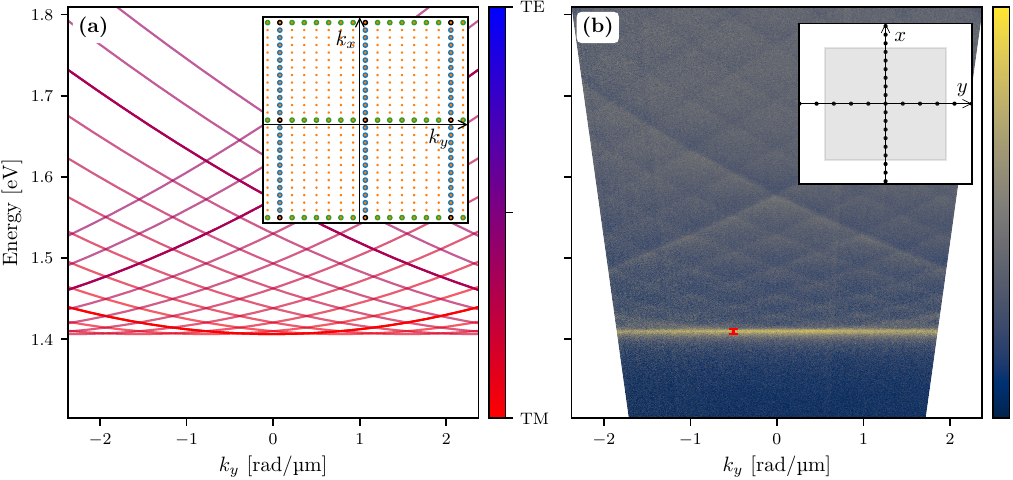}
    \caption{(a) Calculated dispersion and structure factor (inset) of a $\SI{200}{\micro\metre} \times \SI{200}{\micro\metre}$ rectangular chain lattice with $N_x = 13$, $a_x = \SI{580}{\nano\metre}$, $\theta_x = 0^\circ$ and $N_y = 7$, $a_y = \SI{1160}{\nano\metre}$, $\theta_y = 90^\circ$ [see inset in panel (b)] of gold nanocylinders with diameter \SI{130}{\nano\metre} and height \SI{50}{\nano\metre} in an index-symmetric background with the refractive index $n = 1.52$. The calculated dispersion only shows modes with the structure factor $S \geq 0.1$ and with TE-polarization fraction $p_\text{TE} \leq 0.4$. The color scale of the inset has been explained in the main text. The axes have been slightly displaced for clarity. The flat band is formed by the modes on the top and bottom horizontal rows of peaks (green circles). These modes correspond to the diffraction orders with $q_1 = \pm N_x$ or equivalently, $k_x = \pm 2 \pi / a_x$. (b) Experimentally measured extinction spectrum of the same lattice. The extinction spectrum is filtered to show only TM polarization. The linewidth of the flat band indicated by the red bar is \SI{6}{\milli\electronvolt} (\SI{3.7}{\nano\metre}), corresponding to a Q-factor of 235. In the color bar in (b), yellow denotes maximum and blue minimum extinction.}
    \label{f:asymm_cross_exp_vs_sim}
\end{figure*}

The structure factor $S(\mathbf{k})$ for this lattice has been derived in Appendix~\ref{s:S_single_cross} and at the sites of the reciprocal lattice ($\mathbf{k} = q_1 \mathbf{b_1} + q_2 \mathbf{b_2}, \; q_1, q_2 \in \mathbb{Z}$) is
\begin{subequations}
    \label{e:S_asymmetric}
    \begin{empheq}[left={S(\mathbf{k}) = \empheqlbrace}]{align}
        1, \; &\textrm{if}\; q_1 = l_1 N_x \wedge q_2 = l_2 N_y, \label{e:S_asymmetric:a} \\
        \frac{(N_x - 1)^2}{N_u^2}, \; &\textrm{if}\; q_1 = l_1 N_x \wedge q_2 \neq l_2 N_y, \label{e:S_asymmetric:b} \\
        \frac{(N_y - 1)^2}{N_u^2}, \; &\textrm{if}\; q_1 \neq l_1 N_x \wedge q_2 = l_2 N_y, \label{e:S_asymmetric:c} \\ 
        \frac{1}{N_u^2}, \; &\textrm{if}\; q_1 \neq l_1 N_x \wedge q_2 \neq l_2 N_y, \label{e:S_asymmetric:d}
    \end{empheq}
\end{subequations}
where $l_1, l_2 \in \mathbb{Z}$. The structure factor is visualized (for a particular choice of lattice parameters) in the inset of Fig.~\ref{f:asymm_cross_exp_vs_sim}(a). The four cases of Eq.~\eqref{e:S_asymmetric} are apparent here: the global maxima [Eq.~\eqref{e:S_asymmetric:a}] occur at the intersections of the linelike features, and are marked with black in Fig.~\ref{f:asymm_cross_exp_vs_sim}. The secondary maxima Eqs.~\eqref{e:S_asymmetric:b} and~\eqref{e:S_asymmetric:c} form horizontal (vertical) linelike features, marked with green (blue) in Fig.~\ref{f:asymm_cross_exp_vs_sim}. At other sites of the reciprocal lattice [Eq.~\eqref{e:S_asymmetric:d}], the structure factor is negligible; these sites have been marked with orange dots in Fig.~\ref{f:asymm_cross_exp_vs_sim}. The flat band along $k_x = 0$ is formed by modes for which $q_1 = N_x$, i.e., those of Eq.~\eqref{e:S_asymmetric:b} with $l_1 = \pm 1$. For these modes, the structure factor takes a relatively high value ($S \approx 0.4$ for the lattice of Fig.~\ref{f:asymm_cross_exp_vs_sim}). The dense grid of points in the reciprocal space corresponds to the large supercell, while the sparsely distributed points where the structure factor is strongest are given by the (smaller than unit cell) periods of the chains. This is a general feature of chain lattices with multiple lattice sites in the unit cell: in contrast to the rectangular lattice (with only one lattice site in the unit cell) studied in Section~\ref{s:flat_bands_in_ela}, now only certain non-TE modes contribute to the flat band. 

To corroborate these theoretical findings, we measured the extinction spectrum of a 13x7 rectangular chain lattice of gold nanodisks with lattice constants $a_x = \SI{580}{\nano\metre}$ and $a_y = \SI{1160}{\nano\metre}$. The extinction spectrum in TM polarization is shown in Fig.~\ref{f:asymm_cross_exp_vs_sim} along the corresponding calculated band structure: a prominent flat band is visible at \SI{1.41}{\electronvolt}, confirming the prediction that the flat band is TM polarized. Generally, the experimental and calculated dispersions are in good agreement. The flat band is noticeably stronger than the flat band of a rectangular lattice [Fig.~\ref{f:rectangular_lattice_exp_vs_ela}(b)]. We attribute this to the use of the polarization filter, a larger ratio of unit cell size to the lattice constant, and a larger diameter of the nanoparticles. Furthermore, the band is visibly flatter than that of the rectangular array. This is because of the smaller size of the Brillouin zone, i.e., there are more repetitions of the TM modes within the same in-plane wavevector $k_y$ range.

\subsection{Lattices with multiple flat bands}
\label{s:triple_cross}

The results of Section~\ref{s:chain_lattices_main} apply to each constituent chain of a lattice with multiple chains, and therefore we can devise lattice geometries with flat bands at multiple arbitrary energies at the same angle by overlaying two or more parallel chains with appropriate spacings. If we take the orientation of all chains to be the same (to have the flat bands in the same viewing direction), the chains must be offset in space---this is described by the term $\mathbf{\Delta}_j$ in Eq.~\eqref{e:S_chain_lattice}. The offset in space corresponds to a phase shift in the structure factor. For a single chain, the phase factor does not change the structure factor. For multiple chains, the phase factors combine to reduce the structure factor. We illustrate this by an example: consider a lattice of two chains with orientations $\theta_{1,2}$, spacings $a_{1,2}$ consisting of $N_{1,2}$ lattice sites and offset in space by $\mathbf{\Delta}_{1,2}$. Writing out the structure factor [Eq.~\eqref{e:S_chain_lattice}] for this lattice yields
\begin{equation}
    S(\mathbf{k}) = \frac{1}{N_u^2} \left\vert N_1 \varphi_1(\mathbf{k}) S_1(\mathbf{k}) + N_2 \varphi_2(\mathbf{k}) S_2(\mathbf{k}) \right\vert^2.
\end{equation}
From this equation, we draw the following conclusions: First, small offsets $\mathbf{\Delta}_{1,2}$ (so that $\mathbf{k} \cdot \mathbf{\Delta}_{1,2} \approx 0$) reduce the weakening effect. Second, both sublattices can contribute to the structure factor only if $S_1(\mathbf{k})$ and $S_2(\mathbf{k})$ are nonzero (to a sufficient extent) for some value of $\mathbf{k}$. As discussed in Section~\ref{s:chain_lattices_main}, $S_{1,2}(\mathbf{k})$ is non-negligible only near the points given by Eq.~\eqref{e:flat_band_lines}. The criterion of Eq.~\eqref{e:flat_band_lines} can be met for both chains at the same time at individual intersection points of the lines for arbitrary chains [see Fig.~\ref{f:chain_lattice_schematic}(b)]. If, however, both chains have the same orientation and their spacings are the same (or one is an integer multiple of the other), Eq.~\eqref{e:flat_band_lines} is fulfilled by both chains along the lines. 

\begin{figure*}
    \centering
    \includegraphics[width=\linewidth]{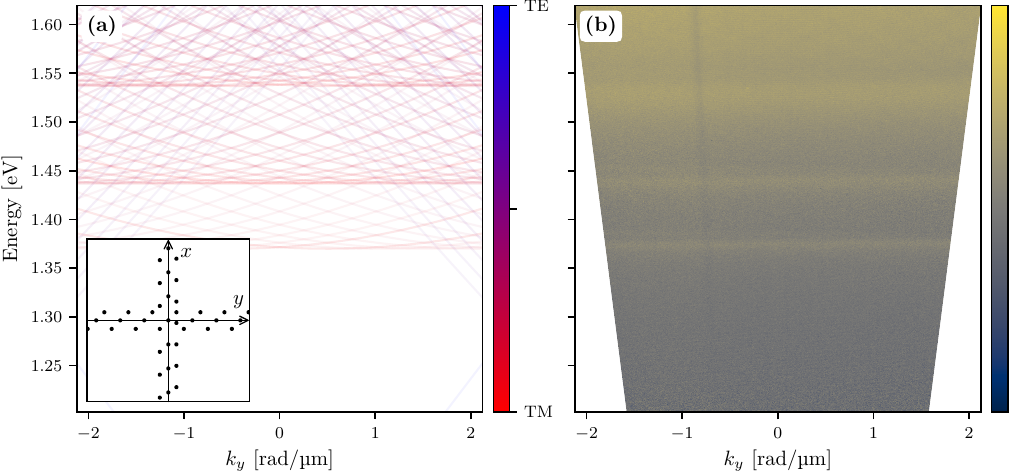}
    \caption{Calculated band structure of a lattice consisting of three rectangular cross lattices (a) and the experimentally measured extinction spectrum of such a geometry realized by nanoparticles (b). Modes with a higher structure factor are plotted in a more opaque (strong) color. The inset shows a part of the unit cell of the lattice. The lattice comprises superimposed 41x21, 43x21, 46x21 rectangular chain lattices with lattice constants $a_x = \SI{592}{\nano\metre}, \; \SI{564}{\nano\metre}, \; \SI{528}{\nano\metre}$ and $a_y = \SI{592}{\nano\metre}$ and offsets $(\Delta x_1, \Delta y_1) = (\SI{0}{\nano\metre}, \SI{0}{\nano\metre}), \; (\Delta x_2, \Delta y_2) = (\SI{200}{\nano\metre}, \SI{200}{\nano\metre}), \; (\Delta x_3, \Delta y_3) = (-\SI{210}{\nano\metre}, -\SI{210}{\nano\metre})$. The size of the unit cell was $\SI{24.3}{\micro\metre} \times \SI{12.4}{\micro\metre}$. The lattice consisted of gold nanocylinders with a diameter of 120 nm and a height of 50 nm in an environment with the refractive index $n = 1.52$. }
    \label{f:triple_cross}
\end{figure*}

We defer further discussion on how to select the offset parameters to Section~\ref{s:double_cross} and for now, to demonstrate that the multiple flat band concept is feasible, designed a lattice consisting of three rectangular chain lattices with different lattice constants sharing a common superperiod. Figure~\ref{f:triple_cross} shows the experimentally measured TM-polarized extinction spectrum and the calculated dispersion of the lattice embedded in an index-symmetric environment. The simulated and experimental data are in good agreement, the three flat bands appear at the predicted energies. The features are relatively weak, with the structure factor taking values $S \approx 0.05$ (maximum would be one) for the modes constituting the flat bands. We conjecture that the flat-band features could be somewhat strengthened by optimizing the offsets of the sublattices, e.g., by maximizing the structure factor of the modes that form the flat bands. However, the maximum value of the structure factor for a chain with $N_j$ lattice sites cannot exceed $N_j^2/N_u^2$, where $N_u$ is the total number of lattice sites in the unit cell. For the lattice of Fig.~\ref{f:triple_cross}, the maximum theoretical values of the structure factor are between 0.10 and 0.13, with the maximum (minimum) obtainable by the chain with the shortest (longest) period. 

\subsection{Lattices with duplicated identical chains}
\label{s:double_cross}

\begin{figure}
    \centering
    \includegraphics[width=\linewidth]{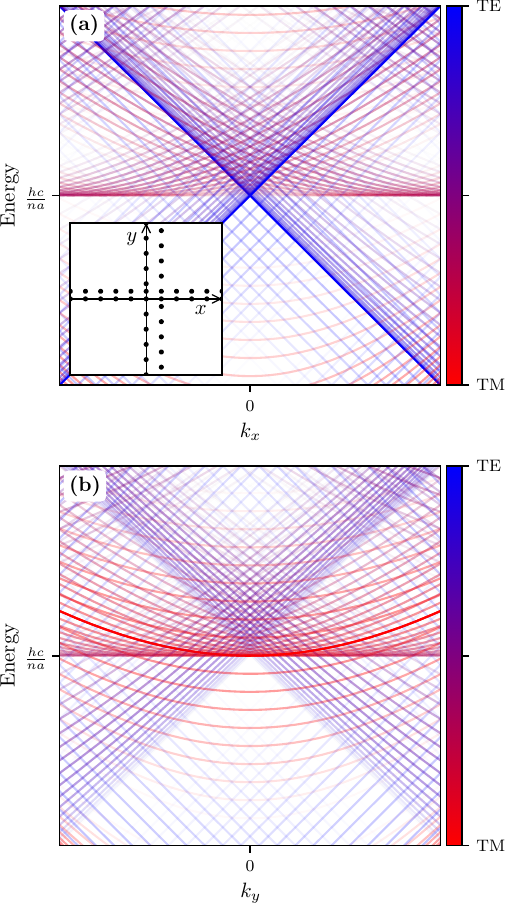}
    \caption{Calculated dispersions of a double $21 \times 21$ rectangular chain lattice with lattice constant $a$ and offsets $\Delta x = a$ and $\Delta y = a/2$ along $k_y = 0$ [panel (a)] and $k_x = 0$ [panel (b)]. Modes with a higher structure factor are plotted in a more opaque (strong) color. The inset of panel (a) shows a part of the unit cell. In determining the polarization of the modes, the plane of incidence is the $xz$ plane in panel (a) and the $yz$ plane in panel (b). }
    \label{f:offsets_effect}
\end{figure}

In the previous Section, we speculated that the choice of the chain offset parameters can affect the structure factor (strength) of particular modes. To better understand the impact of the offset parameters in rectangular chain lattices, we next compare a rectangular $N \times N$ chain lattice with lattice constant $a$ to a lattice that includes a duplicate of the rectangular chain lattice, displaced by $(\Delta x, \Delta y)$ with respect to the other sublattice. The structure factor of the double rectangular chain lattice $S'(\mathbf{k})$ is related to the structure factor of the single $N \times N$ rectangular chain lattice $S(\mathbf{k})$ via (see Appendix~\ref{s:S_double_cross} for the derivation)
\begin{equation}
    % This equation annoyingly runs 6 points too wide of one column, so we just resize it a little bit to make it fit.
    \resizebox{\linewidth}{!}{$
        S'(q_1\mathbf{b}_1{+}q_2\mathbf{b}_2) = \cos^2 \!\left(\frac{\pi}{N a}(q_1 \Delta x{+}q_2 \Delta y)\right) S(q_1\mathbf{b}_1{+}q_2\mathbf{b}_2),
    $}
\end{equation}
where $q_{1,2}$ are integers and $\mathbf{b}_{1,2}$ the reciprocal lattice vectors. From this, it is evident that the offsets can only weaken the modes of the double rectangular chain lattice compared to the single rectangular chain lattice. Considering the flat band along $k_x = 0$ that is formed by the modes corresponding to the diffraction orders $(q_1, q_2) = (N, m), \; m \in \mathbb{Z}$ (as explained in Section~\ref{s:rectangular_cross}), we can maximize the structure factor of the flat-band modes and minimize it for the TE modes by setting $\Delta x = a$, $\Delta y = \frac{a}{2}$. If $N$ is sufficiently large ($m \ll N$), the structure factors of the flat-band modes $(N, m)$ and the (mainly) TE modes $(m, N)$ are approximately
\begin{subequations}
    \begin{align}
        (q_1, q_2) = (N, m): S' &= \cos^2 \left[ \frac{\pi}{N a} \left( N a + m \frac{a}{2} \right) \right]S \approx S, \\
        (q_1, q_2) = (m, N): S' &= \cos^2 \left[ \frac{\pi}{N a} \left( m a + N \frac{a}{2} \right) \right]S \approx 0. 
    \end{align}
\end{subequations}
Physically this means that if one wants to create a strong flat band (high structure factor) from zero up to certain momentum (angle) defined by a certain value of $m$, then $N$ needs to exceed that value. Choosing $\Delta y = \frac{a}{2}$ makes the TE modes near the $\Gamma$ point vanish. Figure~\ref{f:offsets_effect}(a) shows that the above choice of offsets produces a strong flat band along $k_x = 0$. Conversely, along $k_y = 0$ [Fig.~\ref{f:offsets_effect}(b)], the flat band vanishes near the $\Gamma$ point, whereas the TE modes appear strong.

\subsection{Chain lattices with a tilted unit cell}
\label{s:tilted_unit_cell}

Thus far we have studies chain lattices where the orientations of the chains have been parallel to the lattice vectors. Next, we explore lattices where this is no longer the case, starting from a lattice whose unit cell is tilted with respect to the orientation of the chain. In defining such a lattice, there are two issues that must be addressed: selection of the lattice vectors and ensuring that the lattice sites remain within the unit cell. 

In proving that chain lattices give rise to flat bands (Section~\ref{s:chain_lattices_main}), we did not specify the lattice vectors, and thus we can choose the lattice vectors freely, only keeping in mind that the unit cell must be sufficiently larger than the particle spacing to ensure the flatness of the band. To facilitate comparison to the rectangular lattice examined in Section~\ref{s:flat_bands_in_ela}, we study a chain lattice consisting of one chain of $N_u$ lattice sites with spacing $a$ oriented in the $x$ direction, but with the unit cell of the lattice rotated by an angle $\alpha$. The lattice vectors $\mathbf{a}_1$ and $\mathbf{a}_2$ and reciprocal lattice vectors $\mathbf{b}_1$ and $\mathbf{b}_2$ are thus
\begin{align}
    \mathbf{a}_1 &= N_u a \begin{pmatrix}
        \cos \alpha \\ \sin \alpha
    \end{pmatrix},
    &\mathbf{a}_2 = N_u a \begin{pmatrix}
        -\sin \alpha \\ \cos \alpha
    \end{pmatrix}, \nonumber \\
    \mathbf{b}_1 &= \frac{2 \pi}{N_u a}\begin{pmatrix}
        \cos \alpha \\ \sin \alpha
    \end{pmatrix},
    &\mathbf{b}_2 = \frac{2 \pi}{N_u a} \begin{pmatrix}
        -\sin \alpha \\ \cos \alpha
    \end{pmatrix}. \label{e:lattice_vectors_staggered}
\end{align}
We emphasize that it is not necessary to define the lattice vectors using the chain parameters $N_u$ and $a$, it is merely a choice of convenience. 

As the lattice vectors may be chosen freely, it is possible that some of the lattice sites given by Eq.~\eqref{e:line_lattice} end up outside the unit cell spanned by the lattice vectors. To address this, an algorithm to ensure that the lattice sites given by Eq.~\eqref{e:line_lattice} remain in the unit cell while retaining the flat band features for a particular choice of the lattice vectors is presented in Appendix~\ref{s:chain_lattice_generation}. The lattices studied in this work have been generated using this algorithm.

To illustrate the effect of rotating the unit cell with respect to the chain orientation, we study a lattice with $N_u = 5$, $\alpha = \ang{2}$, chain orientation $\theta = \ang{0}$ and lattice vectors given by Eq.~\eqref{e:lattice_vectors_staggered}. The lattice geometry is depicted in the inset of Fig.~\ref{f:lattice_vector_alignment}(a). The algorithm of Appendix~\ref{s:chain_lattice_generation} creates here a chain that appears staggered in the $y$ direction. Despite this, the orientation of the chain is $\theta = \ang{0}$ and it forms a flat band along $k_x = 0$. This is evinced by the calculated band structure of the lattice [Fig.~\ref{f:lattice_vector_alignment}(a)], which shows a relatively flat band near the energy corresponding to the chain spacing. By way of similar reasoning as given in Section~\ref{s:rectangular_cross} and from the structure factor [Fig.~\ref{f:lattice_vector_alignment}(b)], we see that the flat band is formed by the modes $(\pm N_u, m), \, m \in \mathbb{Z}$.

Despite their similar geometries, the band structures of the rectangular lattice [Fig.~\ref{f:rectangular_lattice}(a)] and of the chain lattice with the rotated unit cell [Fig.~\ref{f:lattice_vector_alignment}(a)] are markedly different: the small angle between the chain and one of the lattice vectors lifts the two-fold degeneracy of the flat-band modes leading to a significant distortion of the flat band. The lifting of the degeneracy arises from the fact that the reciprocal lattice [Fig.~\ref{f:lattice_vector_alignment}(b)] is not symmetric about $k_x = 0$. Furthermore, the lifting of the degeneracy caused the linewidth of the flat band to increase. 

We conclude that ensuring that the lines of the peaks of the structure factor given by Eq.~\eqref{e:flat_band_lines} intercept sites of the reciprocal lattice is beneficial for the strength and flatness of the flat band. The location of the reciprocal lattice sites, given by the lattice period (unit cell/supercell size) determines where the structure factor can have non-zero values in the first place, whereas the lines of Eq.~\eqref{e:flat_band_lines}, given by the unit cell structure, determine the locations in the $k_x$-$k_y$ plane where the structure factor can have high values---obviously, for a strong flat band, these two should overlap. 

% Note that the depicted unit cell is indeed the primitive unit cell, because the chain does not lie along a_2.
\begin{figure}[ht]
    \centering
    \includegraphics[width=\linewidth]{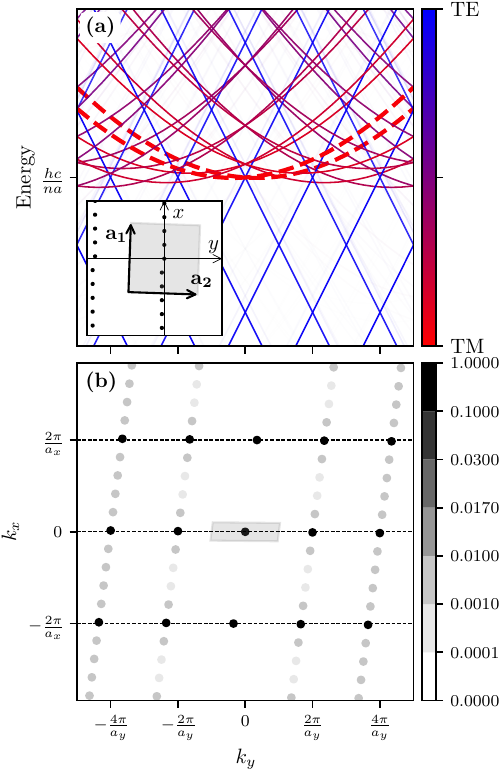}
    \caption{Band structure [panel (a)] and structure factor [panel (b)] of a chain lattice consisting of one five-site chain in the $x$ direction whose lattice vectors $\mathbf{a}_1$ and $\mathbf{a}_2$ are rotated by 2\textdegree{} [see inset in panel (a)]. The two modes highlighted by dashed bold lines in panel (a) are the modes $(\pm 5, 0)$, indicating the lifting of the degeneracy. In panel (b), the shaded area designates the first Brillouin zone and the dashed lines the lines of Eq.~\eqref{e:flat_band_lines}. The color scale of the structure factor has been chosen to highlight that the modes close to the lines of Eq.~\eqref{e:flat_band_lines} are much stronger than any other modes. }
    \label{f:lattice_vector_alignment}
\end{figure}

\subsection{Triangular chain lattices}
\label{s:triangular_chains}

We concluded the previous Section by establishing that ideally, the lines of the peaks of the structure factor given by Eq.~\eqref{e:flat_band_lines} should intersect the sites of the reciprocal lattice. In the rectangular chain lattices (see Section~\ref{s:rectangular_cross}), this was accomplished by aligning the chains with the lattice vectors. Next, by examining a triangular chain lattice, we show that it is possible to add a third chain that aligns with the sites of the reciprocal lattice. 

\begin{figure*}[ht]
    \centering
    \begin{minipage}{0.39\linewidth}
        \includegraphics[width=\linewidth]{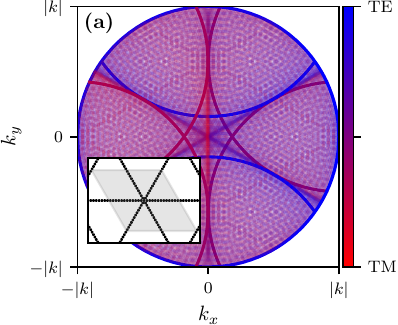}
    \end{minipage}\hfill
    \begin{minipage}{0.39\linewidth}
        \includegraphics[width=\linewidth]{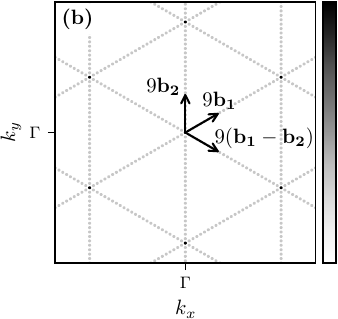}
    \end{minipage}\hfill
    \begin{minipage}{0.2\linewidth}
        \includegraphics[width=\linewidth]{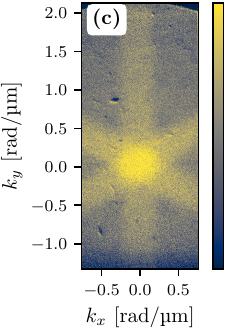}
    \end{minipage}
    \caption{Calculated band structure at the energy corresponding to the lattice constant [panel (a)], structure factor [panel (b)] and extinction spectrum viewed through a filter with central wavelength of 880$\pm 5$ nm [panel (c)] of a triangular chain lattice with $N = 25$ and $a = \SI{570}{\nano\metre}$ in a background with refractive index $n = 1.52$. Inset of panel (a) shows a part of the lattice, with the unit cell shaded in gray. The reciprocal lattice vectors $\mathbf{b}_1$ and $\mathbf{b}_2$ and the difference $\mathbf{b}_1 - \mathbf{b}_2$ are aligned with the lines of the peaks of the structure factor. The strength (color) of the peaks and the length of the vectors are exaggerated for ease of viewing. The extinction spectrum shows three flat bands at the predicted angles. The lattice consisted of gold nanocylinders with a diameter of 150 nm and a height of 50 nm. In the color scale of panel (c), yellow (blue) corresponds to maximum (minimum) extinction. }
    \label{f:triangular_cross}
\end{figure*}

A \emph{triangular chain lattice} consists of three chains with $N$ lattice sites spaced distance $a$ apart at angles 0\textdegree{}, 60\textdegree{}, and 120\textdegree{} to the $x$ axis. The lattice vectors $\mathbf{a}_1$ and $\mathbf{a}_2$ and the reciprocal lattice vectors $\mathbf{b}_1$ and $\mathbf{b}_2$ of the lattice are
\begin{align}
    \mathbf{a}_1 = \begin{pmatrix}
        N a \\
        0
    \end{pmatrix}, \;
    \mathbf{a}_2 = \begin{pmatrix}
        -\frac{1}{2}Na \\
        \frac{\sqrt{3}}{2}Na
    \end{pmatrix}, \nonumber \\
    \mathbf{b}_1 = \begin{pmatrix}
         \frac{2\pi}{Na} \\
        \frac{2\pi}{\sqrt{3} N a}
    \end{pmatrix}, \;
    \mathbf{b}_2 = \begin{pmatrix}
        0 \\
        \frac{4 \pi}{\sqrt{3} Na}
    \end{pmatrix}.
\end{align}
The angles at which the flat bands appear, given by Eq.~\eqref{e:flat_band_angles}, coincide with the vectors $\mathbf{b}_1$, $\mathbf{b}_2$ and $\mathbf{b}_1 - \mathbf{b}_2$ [see Fig.~\ref{f:triangular_cross}(b)]. Furthermore, in a triangular lattice, $|\mathbf{b}_1| = |\mathbf{b}_2| = |\mathbf{b}_1 - \mathbf{b}_2|$, and therefore, all three flat bands have equally many contributing modes. The theoretical predictions are confirmed by empty lattice calculations [Fig.~\ref{f:triangular_cross}(a)] and experimentally [Fig.~\ref{f:triangular_cross}(c)]. 

\subsection{Chain lattices with oblique unit cells}
\label{s:oblique}

\begin{figure}[ht]
    \centering
    \includegraphics[width=\linewidth]{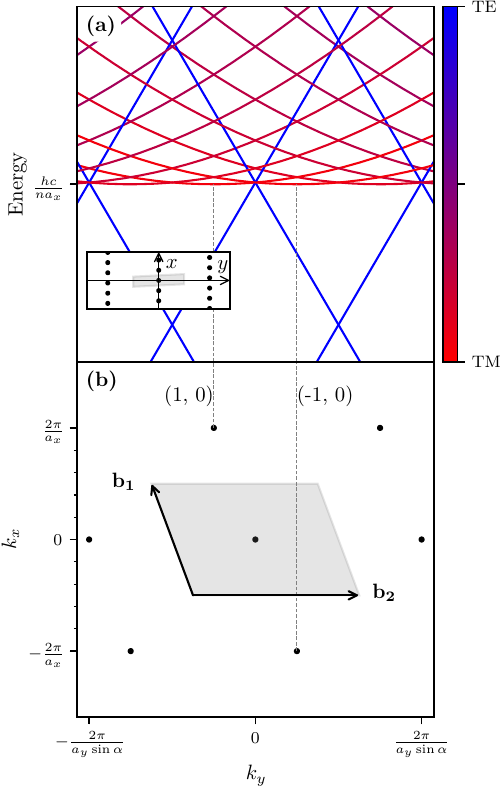}
    \caption{Band structure [panel (a)] and structure factor [panel (b)] of a lattice with lattice vectors given by Eq.~\eqref{e:lattice_vectors} with $a_y = 5 a_x$ and $\alpha = \alpha_0$ [Eq.~\eqref{e:degeneracy_angles}]. The inset of panel (a) shows the real lattice with the unit cell shaded in gray. The shaded area in panel (b) depicts the first Brillouin zone. The flat-band modes closest to the $\Gamma$ point $(\pm 1, 0)$ are indicated in both panels. }
    \label{f:breaking_degeneracy}
\end{figure}

Nonorthogonal lattice vectors lift the degeneracy of the flat-band modes, which can be used to increase the flatness of the flat band by choosing the angle between the lattice vectors $\alpha$ such that it distributes the flat-band modes as evenly as possible. For example, lifting the degeneracy of doubly degenerate modes, and then placing them evenly, improves the flatness of the resulting band effectively by a factor of two. Moreover, one of the lattice vectors is aligned with the chain orientation to ensure that the degeneracy is lifted only by translating the modes with respect to the wavevector $k_y$ and not the energy (as was the case for the lattice of Fig.~\ref{f:lattice_vector_alignment}). To illustrate, we consider a rectangular lattice with lattice vectors and reciprocal lattice vectors
\begin{align}
    \mathbf{a}_1 &= \begin{pmatrix}
        a_x \\ 0
    \end{pmatrix},
    &\mathbf{a}_2 = a_y \begin{pmatrix}
        \cos \alpha \\ \sin \alpha
    \end{pmatrix}, \nonumber \\
    \mathbf{b}_1 &= \frac{2 \pi}{a_x}\begin{pmatrix}
        1 \\ -\cot \alpha
    \end{pmatrix},
    &\mathbf{b}_2 = \frac{2 \pi}{a_y} \begin{pmatrix}
        0 \\ \frac{1}{\sin \alpha}
    \end{pmatrix}, \label{e:lattice_vectors}
\end{align}
where we assume $a_y > a_x$ so that there is a flat band along the line $k_x = 0$, i.e., the (first) flat band is formed by the modes $(\pm 1, q_2), \, q_2 \in \mathbb{Z}$. Next, to distribute the modes that constitute the flat band maximally equally, we require that the $k_y$ coordinates of the modes related to diffraction orders $(\pm 1, q_2)$, for some value of $q_2$ (we choose $q_2 = 0$) are separated by a half-integer multiple of the width of the Brillouin zone ($y$ component of $\mathbf{b}_2$) (the $k_y$ coordinates of the modes are given by the $y$ component of the reciprocal vector $ \mathbf{b}_1$, as is intuitive from Fig.~\ref{f:breaking_degeneracy}). This yields the following condition for the angle of the lattice vectors $\alpha$:
\begin{align}
    2(b_1)_y &= \left(j + \frac{1}{2}\right) (b_2)_y \nonumber \\
    \Rightarrow \alpha_j &= \arccos \left[- \left(j + \frac{1}{2}\right)\frac{a_x}{2 a_y} \right], \label{e:degeneracy_angles}
\end{align}
where $j$ is an integer in the interval $\left[-\frac{2 a_y}{a_x} - \frac{1}{2}, \frac{2 a_y}{a_x} - \frac{1}{2}\right]$. Figure~\ref{f:breaking_degeneracy} shows the band structure of a lattice with $a_y = 5 a_x$ and the angle of the lattice vectors $\alpha = \alpha_0 \approx \ang{87.1}$. Compared to the corresponding rectangular lattice (Fig.~\ref{f:rectangular_lattice}), there are more nondegenerate modes in a given $k_y$ interval, and thus the band is flatter.

\subsection{Multichain lattices, flat bands with a finite angular extent}
\label{s:multichain_lattices}

In Section~\ref{s:double_cross}, we established that adding copies of a chain with the same spacing and angle can only reduce the structure factor. We show here that this provides means to design lattices with flat bands with a desired, finite angular extent. To this end, we next consider a lattice whose unit cell is an $a_x \times a_y$ rectangle that includes $N$ lattice sites at positions $\{ (0, \Delta_n) \}, \; n \in [0, \ldots, N - 1]$. We take $a_y > a_x$ so that there is a flat band along $k_x = 0$. 

Using Eq.~\eqref{e:def_S}, the structure factor of this lattice can be evaluated to yield 
\begin{equation}
    S(\mathbf{k}) = \frac{1}{N^2}\left\vert \sum_{n=0}^{N-1} e^{i k_y \Delta_n}\right\vert^2.
\end{equation}
We cannot analytically solve the positions $\{\Delta_n\}$ that yield a given structure factor. To be able to connect the angular extent of the flat band to the design parameters, we select $\Delta_n = \Delta_y n$, where $\Delta_y$ is a constant spacing parameter. With this choice, the structure factor $S(\mathbf{k})$ becomes
\begin{equation}
    S(\mathbf{k}) = \frac{1}{N^2} \frac{1 - \cos \left(\Delta_y k_y N\right)}{1 - \cos \left(\Delta_y k_y \right)} \approx 1 - \frac{1}{12}(N^2 -1)(\Delta_y k_y)^2.
\end{equation}
The latter approximate value is valid for small values of the arguments of the cosines. 
The structure factor has a maximum at $k_y = 0$, $S(k_y = 0) = 1$, and its minimal value is zero. Thus from the above Taylor series expression at $k_y=0$, we can solve for the value of $k_y$ corresponding to half-width at half-maximum $k_\textrm{HWHM}$ ($S(k_\textrm{HWHM})=0.5$):
\begin{equation}
    k_\textrm{HWHM} = \frac{\sqrt{6}}{\Delta_y \sqrt{N^2-1}}.
\end{equation}
This value approximately characterizes the extent of the flat band in momentum. Intuitively one can understand this effect as destructive interference caused by the multiple particles in the unit cell. However, for small $k_y$ the corresponding length scale (effective wavelength) is large compared to the spacing between the particles, and even the total length $N \Delta_y$, thus this effectively subwavelength feature does not cause a major effect, and one still has a flat band over small values of $k_y$.

At the energy of the flat band [which is determined by $a_x$ per Eq.~\eqref{e:flat_band_energy}], the corresponding angle $\theta_\textrm{HA}$ is 
\begin{equation}
    \theta_\textrm{HA} = \arcsin \left(\frac{k_\textrm{HWHM}}{k_0}\right) = \arcsin \left(\frac{a_x }{2 \pi \Delta_y}\sqrt{\frac{6}{N^2-1}}\right).
\end{equation}
Finally, to design a lattice with a given angular range $\theta_\textrm{HA}$, we solve for the $y$-offset magnitude $\Delta_y$:
\begin{equation}
    \label{e:delta_y}
    \Delta_y = \frac{a_x}{\pi \sin \theta_\textrm{HA}}\sqrt{\frac{3}{2(N^2-1)}}.
\end{equation}
Notably, the lattice constant $a_y$ does not directly appear in the expression for $\Delta_y$. However, it controls how many modes appear within the angular range determined by the half-angle $\theta_\textrm{HA}$.

\begin{figure}[ht]
    \centering
    \includegraphics[width=\linewidth]{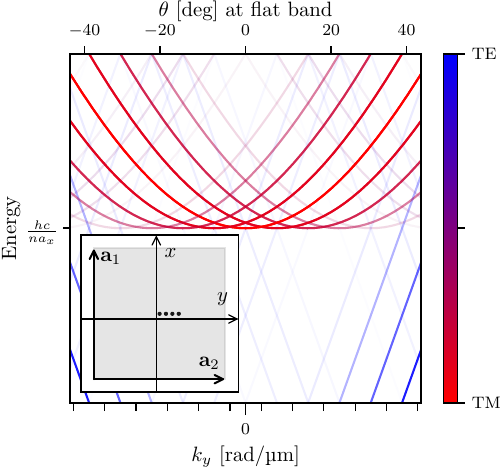}
    \caption{Band structure of a lattice designed to have a flat band between $\pm \ang{15}$. The inset shows the unit cell (in gray) of the lattice, which has four particles, with spacing of $\Delta_y \approx 0.389 a_x$ [calculated using Eq.~\eqref{e:delta_y}]. Note that the inset is not to scale, $a_y = | \mathbf{a}_2| = 8 a_x$ ($a_x = |\mathbf{a}_1|$); this value of $a_y$ was selected to best visualize the weakening of the flat band at greater angles. The axes in the inset have been slightly displaced for ease of viewing, the leftmost lattice site is actually at the origin. The spacing of the minor ticks of the $k_y$ axis corresponds to the width of the Brillouin zone.}
    \label{f:finite_angle_flat_band}
\end{figure}

To verify the above predictions, Fig.~\ref{f:finite_angle_flat_band} shows the band structure of a lattice designed to have a flat band extending to $\theta_\textrm{HA} = \ang{15}$. As expected, the modes whose vertices are at $k_y$ values corresponding to angles greater than $\ang{15}$ appear weak. 

% An alternative way to visualize this would be to only plot the modes for which S >= 0.5. But maybe this is better, to clearly indicate that the scattering strength decreases as a function of angle.

\subsection{Summary of the results}
\label{s:results_summary}
% This could of course be under Conclusions and discussion, too.

We have studied several examples of chain lattices, and now summarize the results as the following design rules for creating lattices exhibiting flat bands:
\begin{enumerate}
    \item A 1D chain of lattice sites with constant spacing $a_j$ subtending an angle $\theta_j$ with the $x$ axis gives rise to a flat band that is observable perpendicular to the angle $\theta_j$ and at wavelength $n a_j$, where $n$ is the refractive index of the background medium. If multiple chains with different spacing or orientation are overlaid in the same lattice, multiple flat bands appear.
    \item The contribution of the $j$th chain [captured in the partial sum $S_j$ of Eq.~\eqref{e:S_chain_lattice}] to the structure factor $S$ is proportional to the number of lattice sites in the chain $N_j^2$, while the structure factor is normalized by the number of lattice sites in the unit cell $N_u^{-2}$. Thus, the distribution of the lattice sites to different chains can be used to control the relative strengths of the modes pertaining to each chain.
    \item If multiple copies of a chain with the same periodicity, but spatially offset by $\mathbf{\Delta}_j$, are added into a lattice, the effect of the offsets $\mathbf{\Delta}_j$ is to weaken the modes at the crossings of the $k$-space lines originating from the chains $j$ and $j' \neq j$, i.e., modes common to the chains $j$ and $j'$; this effect is captured in the term $\varphi_j(\mathbf{k})$ in Eq.~\eqref{e:S_chain_lattice}. The weakening effect is the most severe for parallel chains because it applies to all modes.
    \item In the optimal case, the lines of the peaks of the structure factor defined by Eq.~\eqref{e:flat_band_lines} intersect the sites of the reciprocal lattice. The effect of the misalignment between the lines of the peaks of the structure factor and the sites of the reciprocal lattice on the quality of the flat band can be reduced by increasing the density of the modes, i.e., by increasing the size of the unit cell in real space.
    \item Nonorthogonal lattice vectors can be used to lift the degeneracy of the modes that form the flat band. This can be used to increase the flatness of the band by selecting the angle of the lattice vectors using Eq.~\eqref{e:degeneracy_angles}.
    \item Flat bands arising from chain lattices naturally extend over all angles. Lattices yielding flat bands with a finite angular range can be designed by adding multiple copies of the same chain, offset by a constant step selected using Eq.~\eqref{e:delta_y}.
\end{enumerate}

\section{Conclusions and discussion}
\label{s:conclusions}

We have theoretically formulated and experimentally demonstrated a strategy to create flat bands of diffractive origin by lattices where the width of the unit cell in one direction is large compared to the lattice site spacing in the perpendicular direction. These flat bands naturally extend over all angles, but can be tailored to cover only a finite angular range if desired. The flat bands are dominantly TM polarized at small angles. We introduced a category of lattice geometries, the chain lattices, and showed that, using equispaced 1D chains of lattice sites as building blocks, superlattice geometries can be engineered to host flat bands at energies and angles that are determined by the geometry of the sublattice chain via simple analytical relations. Intuitively, one can conclude that a single 1D chain exhibits a flat band in the direction perpendicular to the chain because of lack of structure in that direction, while the periodicity of the chain in the other direction allows selecting a specified, narrow value for the energy of the band. The flat-band design method was demonstrated by several representative examples, and in each case the theoretical predictions matched well with experimental Fourier spectroscopy data from lattices of gold nanocylinders fabricated by electron-beam lithography. Because of their purely diffractive origin, we expect the flat bands to also appear in various other lattices of scatterers, e.g., arrays of dielectric nanoparticles where the absence of ohmic losses would help achieving even higher-quality modes. 

1D chains of nanoparticles have been investigated in Ref.~\cite{Rekola2018}, and the authors reported flat-band lasing. Our work complements this observation by providing a theoretical explanation for the appearance of the flat band. Superlattice geometries (where the particle spacing is small compared to the size of the unit cell) have previously been investigated~\cite{Wang2015, Wang2017, Wang2023}, but in those studies, the square geometry of the sublattice excludes flat bands. While asymmetric, two-dimensional rectangular arrays have previously been employed to realize flat bands~\cite{Do2025,Eyvazi2025}, in those systems the band flattening arises from coupling of guided modes or refractive index contrast with the environment. In contrast, the flat bands reported here originate from diffraction. Moreover, chain lattices do not require fine-tuning of the rectangular aspect ratio to achieve band flattening, relaxing the geometric constraints inherent to previously reported implementations. In another realization, a lattice of crossing dielectric waveguides---similar in appearance to the rectangular chain lattices introduced in this work---was observed to feature bidirectional flat bands~\cite{Sun2025-2}, however the underlying physical mechanism is based on combining a metasurface--waveguide with features of tight-binding systems, which is very different from our diffraction-originated flat bands. 

Previous realizations of flat bands in photonic systems have relied on (i) limiting the range of coupling using arrays of parallel waveguides~\cite{GuzmanSilva2014, Vicencio2015, Xia2018, Cantillano2018, Song2023, mukherjee2015observation}, photonic moir\'e lattices~\cite{Yan2025, Wang2020, Jing2025, Yi2022, Ning2023}, coupled metallic resonators~\cite{Yang2024, Kajiwara2016, Nakata2012} or short-lived exciton-polaritons~\cite{jacqmin2014direct, baboux2016bosonic, harder2021kagome}, (ii) using gratings to couple free space modes to waveguided modes~\cite{amedalor2023high, nguyen2018symmetry, choi2024nonlocal, munley2023visible, Le2024, Choi2025, Eyvazi2025, Choi2025-2, Sun2025-2}, (iii) using strain to flatten the Landau levels~\cite{Barczyk2024}, (iv) exploiting the rotational degrees of freedom of the constituent elements of the unit cell~\cite{Xia2025}, or (v) band folding arising from large superlattice periodicity and disorder-induced qBICs~\cite{Qin2025}. The flat bands investigated here arise purely from diffraction, through the superposition of TM modes that are relatively flat near their vertices. One of the advantages of this approach is that the flat bands naturally extend over all angles, whereas the flat bands arising from coupling to waveguide modes, for example, have a limited angular range, and the moir\'e bands are only approximately flat and only for precise conditions. 

The SLRs of arrays of plasmonic nanoparticles have been used to enhance and control the emission of organic dye molecules~\cite{Vecchi2009}, OLEDs~\cite{Auer-Berger2017-1, Auer-Berger2017-2, Auer-Berger2022}, and LEDs~\cite{Abdelkhalik2023, Lozano2013}. For these applications---and for other potential applications such as increasing absorption in photodetectors or photovoltaics---the chromatic dispersion of SLRs of simple lattice geometries is usually an unwanted effect. The lattice geometries presented in this work offer means to enhance emission or absorption at a narrow energy range over all or desired angles. In particular, lattices featuring flat bands at multiple target energies are interesting for white-light generation or absorption. Furthermore, the flat bands being linearly TM polarized may be beneficial to applications requiring polarized light. Lattices yielding flat bands with a finite, tunable angular range may be useful, e.g., in OLEDs, where the range could be matched with the range of angles ($k_\parallel$ values) that is allowed to outcouple before total internal reflection at higher angles. 

In designing the chain lattices, there is an ineluctable tradeoff: while a larger unit cell produces a flatter band, it also leads to a reduced filling factor, thus weakening the light--matter interaction. Despite this, the experimental data presented here show that designing lattice geometries with visibly flat bands is feasible. Furthermore, we showed that the band flatness can be increased by a suitable choice of the angle of the lattice vectors. Flat-band lasing has been reported from photonic moir\'e lattices~\cite{harder2021kagome, Mao2021, Luan2023} and metasurface-on-a-waveguide structures~\cite{Eyvazi2025}. The lattice geometries proposed here offer an exciting platform for plasmonic flat-band lasing and amplified spontaneous emission; experimental results on those topics are presented in Ref.~\cite{Heilmann2025}. Other potential applications include slow-light devices and enhancing nonlinear optical processes, such as harmonic generation. As our strategy for creating flat bands relies on simple generic principles expressed by analytical formulas, in contrast to numerical approaches, it can be easily adapted to different systems and applications.

\section*{Competing interests}
R. H., J. L., A. J. J. D, and P. T. declare that they are inventors in a filed PCT (PCT/FI2026/050087) patent application related to the work described in this manuscript. The remaining authors declare no competing interests.

\begin{acknowledgments}
Funded by the European Union. Views and opinions expressed are however those of the author(s) only and do not necessarily reflect those of the European Union or the European Innovation Council and the SMEs Executive Agency (EISMEA). Neither the European Union nor the granting authority can be held responsible for them (SCOLED, Grant Agreement No. 101098813). This work is part of the Finnish Centre of Excellence in Quantum Materials (QMAT).

The research was performed at the Micronova Nanofabrication Centre of Aalto University.
\end{acknowledgments}

\appendix

\section{Sample fabrication}
\label{s:sample_fabrication}
The nanoparticle arrays are fabricated using electron-beam lithography (EBL). For this, a resist (350~$\mu$l PMMA A4) is spin-coated onto a 25~mm x 75~mm x 1~mm large borosilicate substrate for 1 minute at 3000~rpm, followed by a 2-minute bake at 175$^{\circ}$C on a hot plate. To create a conductive and reflective surface for the lithography process, a 15~nm-thick layer of aluminum is evaporated onto the sample using electron-beam evaporation. After exposure at the EBL, the aluminum layer is etched for 1 min in a 1:1 mixture of AZ351B-developer and DI-water, followed by a 30~s bath in pure DI-water. The resist is developed for 15 s in a 3:1 mixture of isopropanol:MIBK (Methyl isobutyl ketone) followed by a 30~s bath in pure isopropanol. Following the development, 2~nm of titanium and 50~nm of gold are evaporated onto the sample using electron beam evaporation. The excess material is removed in a lift-off process by immersing the sample overnight in acetone.

For the measurements, the arrays are immersed in index-matching oil, and the sample is sealed with a 25~mm x 25~mm x 0.1~mm borosilicate cover slip, to ensure that the arrays are in an index-symmetric environment with a refractive index of 1.52.

\section{Experimental setup}
\label{s:experimental_setup}
The sample is illuminated with a transmission light source producing diffused white light. The objective (NA~0.3) is used to image the arrays. The back focal plane of the objective is guided to the entrance slit of the spectrometer. The back focal plane of the objective corresponds to the momentum space and hence contains information about the angular properties of the light transmitted through the sample. In the spectrometer, the white light is separated into different wavelengths with a grating and guided to a charge-coupled device (CCD) camera, which records the angle- and wavelength-resolved images. Optionally, a polarizer can be added to the path. A real-space camera is used to record the real-space of the sample and to bring the sample into the focal point of the objective. A real-space iris is used to restrict the light to only the area that is covered by a nanoparticle array. The transmission through the sample $T_\textrm{array}$ is calculated in a post-processing step by
\begin{equation}
    T_\textrm{array}=\frac{I_\textrm{array}-I_\textrm{background}}{I_\textrm{reference}-I_\textrm{background}},
\end{equation}

where $I_\textrm{array}$ is the intensity recorded in the CCD camera of the spectrometer through the array, $I_\textrm{background}$ is the intensity of the background radiation which is collected by blocking the direct beam path to the spectrometer, and $I_\textrm{reference}$ is the intensity through the sample with no array present, i.e., through the substrate, index-matching oil and cover slip layers. Extinction, shown in the pictures of the main text, is obtained as $1 - T_{\textrm{array}}$.

\begin{figure}
    \centering
    \includegraphics[width=0.5\linewidth]{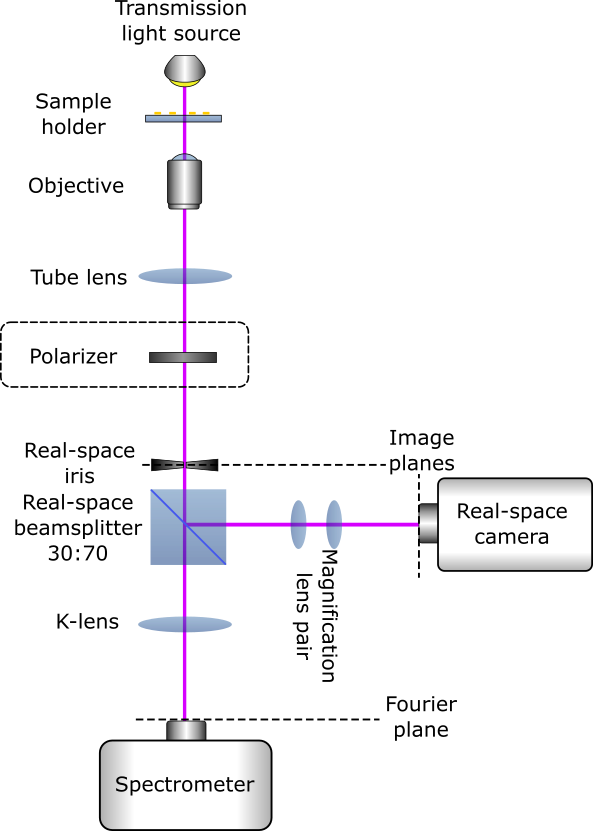}
    \caption{Schematic of the Fourier imaging setup. }
    \label{f:experimental_setup}
\end{figure}

\section{Structure factor of a rectangular chain lattice}
\label{s:S_single_cross}

The positions of the lattice sites in an $N_x \times N_y$ rectangular chain lattice with the lattice constants $a_x$ and $a_y$ are
\begin{equation}
    \biggl\{(n a_x, 0)\biggr\}_{n=0}^{N_x-1} \, \bigcup \, \biggl\{(0, n a_y)\biggr\}_{n=0}^{N_y-1},
\end{equation}
where $\bigcup$ indicates a union of two sets. There are $N_u = N_x + N_y - 1$ lattice sites in the unit cell; the $-1$ appears to avoid counting the lattice site at the origin twice. The lattice vectors and the reciprocal lattice vectors are given by Eq.~\eqref{e:lattice_vectors_rect_chain}. Inserting the positions of the lattice sites to the definition of the structure factor [Eq.~\eqref{e:def_S}] yields
\begin{equation}
    \label{e:S_cross}
    S(\mathbf{k}) = \frac{1}{N_u^2} \left|-1 + \sum_{n=0}^{N_x-1} e^{i k_x n a_x} + \sum_{n=0}^{N_y-1} e^{i k_y n a_y} \right|^2,
\end{equation}
where the $-1$ again appears to avoid double-counting the lattice site at the origin. Per Section~\ref{s:flat_bands_in_ela}, the structure factor can have non-zero values only at the sites of the reciprocal lattice, i.e., $\mathbf{k} = q_1 \mathbf{b_1} + q_2 \mathbf{b_2}$ with $q_1, q_2 \in \mathbb{Z}$. The geometric sums in Eq.~\eqref{e:S_cross} then evaluate to
\begin{align}
    &\sum_{n=0}^{N_x-1} e^{i k_x n a_x} = \sum_{n=0}^{N_x-1} e^{2 \pi i q_1 \frac{n}{N_x}} \nonumber \\
    &= \begin{cases}
        N_x, \textrm{ if } q_1 = l_1 N_x, \, l_1 \in \mathbb{Z}, \\
        0, \textrm{ else.}
    \end{cases}
\end{align}
Evaluating the other sum in a similar manner, we obtain the expression for the structure factor
\begin{equation}
    S(q_1 \mathbf{b_1} + q_2 \mathbf{b_2}) = \begin{cases}
        1, &\textrm{if}\; q_1 = l_1 N_x \wedge q_2 = l_2 N_y, \\
        \frac{(N_x - 1)^2}{N_u^2}, \; &\textrm{if}\; q_1 = l_1 N_x \wedge q_2 \neq l_2 N_y, \\
        \frac{(N_y - 1)^2}{N_u^2}, \; &\textrm{if}\; q_1 \neq l_1 N_x \wedge q_2 = l_2 N_y, \\ 
        \frac{1}{N_u^2}, \; &\textrm{if}\; q_1 \neq l_1 N_x \wedge q_2 \neq l_2 N_y,
    \end{cases}
\end{equation}
where $l_1, l_2 \in \mathbb{Z}$.

\section{Structure factor of a symmetric double rectangular chain lattice}
\label{s:S_double_cross}
We consider a lattice with the (reciprocal) lattice vectors
\begin{align}
    \mathbf{a_1} = \begin{pmatrix}
        N a \\
        0
    \end{pmatrix} , \;
    \mathbf{a_2} = \begin{pmatrix}
        0 \\
        N a
    \end{pmatrix}, \nonumber \\
    \mathbf{b_1} = \begin{pmatrix}
        \frac{2 \pi}{N a} \\
        0
    \end{pmatrix} , \;
    \mathbf{b_2} = \begin{pmatrix}
        0 \\
        \frac{2\pi}{N a}
    \end{pmatrix},
\end{align}
with two $N \times N$ rectangular chain lattices with relative displacement $(\Delta x, \Delta y)$ [see inset of Fig.~\ref{f:offsets_effect}(a) in the main text]. The unit cell has $N_u' = 4N - 2$ lattice sites in the unit cell (-2 is to avoid double-counting the sites at the centers of both crosses). To highlight the effect of adding the second pair of chains and the choice of the offset parameters $(\Delta x, \Delta y)$, we compare the structure factor of the double rectangular chain lattice to that of a single rectangular chain lattice, which was calculated in Appendix~\ref{s:S_single_cross}. We denote the structure factor of the single rectangular chain lattice by $S(\mathbf{k})$ and that of the double rectangular chain lattice by $S'(\mathbf{k})$. Noting that the double rectangular chain lattice has twice as many lattice sites in the unit cell as the single rectangular chain lattice ($N_u' = 2 N_u$), we compute the structure factor $S'(\mathbf{k})$ by inserting the positions of the lattice sites in the definition of the structure factor [Eq.~\eqref{e:def_S}]:
\begin{align}
    S'(\mathbf{k}) &= \frac{1}{N_u'^2} \left| \sum_{n = 0}^{N-1} \bigl[e^{i k_x n a} + e^{i k_y n a} + \right.\nonumber \\
    &\left.\overbrace{e^{i (k_x \Delta x + k_y \Delta y)}}^{\equiv \varphi(\mathbf{k})} \left(e^{i k_x n a} + e^{i k_y n a}\right)\bigr] - 2 \right|^2 \nonumber \\
    &= \frac{1}{4N_u^2} \left| [1 + \varphi(\mathbf{k})] \sum_{n = 0}^{N-1} \left[e^{i k_x n a} + e^{i k_y n a}\right] -2\right|^2 \nonumber \\
    &= \frac{|1 + \varphi(\mathbf{k})|^2}{4} \frac{1}{N_u^2} \left| \sum_{n = 0}^{N-1} \left[e^{i k_x n a} + e^{i k_y n a}\right] -2\right|^2 \nonumber \\
    &= \frac{|1 + \varphi(\mathbf{k})|^2}{4} S(\mathbf{k}).
\end{align}
Recalling that $S$ can take nonzero values only at the sites of the reciprocal lattice $\mathbf{k} = q_1 \mathbf{b_1} + q_2 \mathbf{b_2}$, we can write 
\begin{equation}
    \frac{|1 + \varphi(q_1 \mathbf{b_1} + q_2 \mathbf{b_2})|^2}{4} = \cos^2 \left[\frac{\pi}{N a}(q_1 \Delta x + q_2 \Delta y)\right].
\end{equation}

\section{Generation of chain lattices}
\label{s:chain_lattice_generation}

In generating the coordinates of the lattice sites per Eq.~\eqref{e:line_lattice}, some of the lattice sites may end up outside the unit cell defined by the lattice vectors $\mathbf{a}_{1,2}$. Any lattice sites outside the unit cell must be translated back into the unit cell in a manner that does not affect their contribution to the structure factor. To this end, we note that the structure factor [Eq.~\eqref{e:def_S}] is invariant under a translation of a lattice site by an integer multiple of the lattice vectors. Thus, the steps of the algorithm are simply:
\begin{enumerate}
    \item Generate the coordinates in the global $xy$-coordinate system using Eq.~\eqref{e:line_lattice}.
    \item Write the coordinates in the basis of the lattice vectors $\mathbf{a}_{1,2}$.
    \item Discard the integer part of the coordinates, keeping only the fractional part. 
\end{enumerate}

For step 2, we denote the coordinates of the $j$th lattice site in the basis of the lattice vectors by $\mathbf{\tilde c}_j$. They are given by the coordinate transformation
\begin{equation}
    \begin{pmatrix}
        \tilde c_{j1} \\
        \tilde c_{j2}
    \end{pmatrix} = A^{-1} \left[
    n \begin{pmatrix}
        c_{jx} \\
        c_{jy}
    \end{pmatrix} + \begin{pmatrix}
        \Delta x_j \\
        \Delta y_j
    \end{pmatrix}\right], \label{e:line_lattice_coordinate_transformation}
\end{equation}
where $A$ is the matrix whose columns are the lattice vectors. Because $\mathbf{a_i} \cdot \mathbf{b_j} = 2 \pi \delta_{ij}$, where $\mathbf{b}_{1,2}$ are the reciprocal lattice vectors, it follows that 
\begin{equation}
    A^{-1} = \frac{1}{2 \pi} B^\text{T} = \frac{1}{2 \pi} \begin{pmatrix}
        b_{1x} & b_{1y} \\
        b_{2x} & b_{2y}
    \end{pmatrix},
\end{equation}
where $B$ is a matrix whose columns are the reciprocal lattice vectors. 

\begin{figure}
    \centering
    \includegraphics[width=\linewidth]{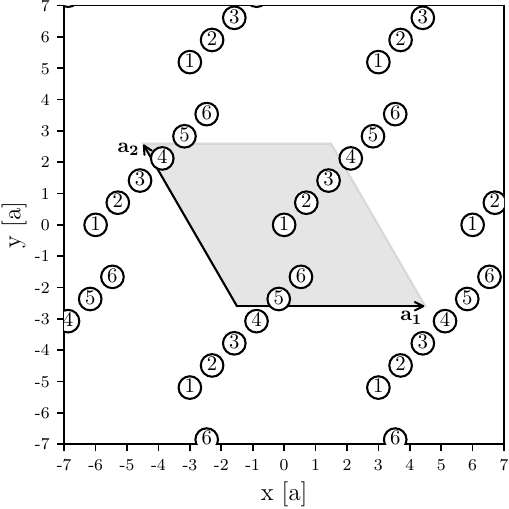}
    \caption{Illustration of the chain lattice generation algorithm. The depicted lattice consists of one chain with $N = 6$ sites, spacing of $a$, and orientation $\theta = \ang{45}$. The lattice vectors are $\mathbf{a}_1 = (N a, \; 0)^\mathrm{T}$ and $\mathbf{a}_2 = N a(\cos \ang{120}, \; \sin \ang{120})^\mathrm{T}$. All sites marked with the same number are related to one another by a translation of an integer multiple of the lattice vectors.}
    \label{f:chain_lattice_generation}
\end{figure}

Step 3 is accomplished by the replacement $\mathbf{\tilde c_j} \rightarrow \mathbf{\tilde c_j} - \lfloor \mathbf{\tilde c_j} \rfloor$, where $\lfloor \, \rfloor$ denotes element-wise rounding down. Here, "rounding down" for negative numbers means rounding away from zero (e.g.,  $\lfloor-2.2\rfloor = -3$)---this choice ensures that the coordinates are always in the range $[0, 1)$. Figure \ref{f:chain_lattice_generation} illustrates the output of this algorithm. 

% PRX requires all references to be in the main document, even if they are only referenced in the supplementary (https://journals.aps.org/authors/supplemental-material-instructions). To have them included in the bibliography, we need to \nocite{} the entries that have not been cited in the main document.
% As we did not include the PL enhancement in the end, the supplementary no longer references the COMSOL sources. If that is reintroduced at some point, the following line should be uncommented.
%\nocite{Comsol, ComsolModeAnalysis, ComsolModeTracking}

% The publicly available dataset needs to be referenced in the paper.
\nocite{dataset}

\bibliography{flat_bands,plasmonics,oleds,supplemental_material,miscellaneous}

\end{document}